%% file: SFRenorm.tex
\documentclass[a4paper,11pt,twoside]{article}
\usepackage{amsmath,amssymb,amsthm}
\usepackage[ps,dvips,arc,frame]{xy}
\usepackage[dvips]{graphicx}

\input{macrodef}

\begin{document}

\input{titlepage}
\tableofcontents
\newpage
\input{intro}
\input{cellular}
\input{rengrp}
\input{diag}
\input{models1}
\input{models2}
\input{conclude}
\input{acknowledge}

\bibliographystyle{amsordx}
\bibliography{stdrefs}
\end{document}

%% file: macrodef.tex
\newcommand{\im}{\mathrm{i}}  
\newcommand{\defeq}{:=}

\newcommand{\xd}{\mathrm{d}}  
\newcommand{\xD}{\mathcal{D}}  

\DeclareMathOperator{\tr}{tr}

\newcommand{\pf}{\mathcal{Z}}

\newcommand{\pfx}[1]{\mathcal{Z}_{\text{#1}}}
\newcommand{\pfmx}[1]{\tilde{\mathcal{Z}}_{\text{#1}}}
\newcommand{\cdec}{\mathcal{C}}
\newcommand{\dskel}{\mathcal{K}}
\newcommand{\SU}{\mathrm{SU}}
\newcommand{\Spin}{\mathrm{Spin}}

\theoremstyle{plain}
\newtheorem{prop}{Proposition}[section]

\newtheorem{dfn}[prop]{Definition}
\newtheorem{con}[prop]{Conjecture}
\theoremstyle{definition}

\newcommand{\rxy}[1]{{\begin{xy} 0;<1mm,0mm>:<0mm,1mm>::0;0,#1
\end{xy}}}

%% file: titlepage.tex
\begin{titlepage}
\title{\textbf{Renormalization of Discrete Models without Background}}
\author{Robert Oeckl\footnote{email: oeckl@cpt.univ-mrs.fr}\\ \\
Centre de Physique Th\'eorique,
CNRS Luminy,\\
13288 Marseille cedex 9, France}
\date{CPT-2002/P.4461\\
22 January 2003 (v2)}

\maketitle

\vspace{\stretch{1}}

\begin{abstract}
\input{abstract}
\end{abstract}

\vspace{\stretch{1}}
\end{titlepage}

%% file: abstract.tex
Conventional renormalization methods in statistical physics and
lattice quantum field theory assume a flat metric background.
We outline here a generalization of such methods to models on
discretized spaces without metric background. Cellular decompositions
play the role of discretizations. The
group of scale transformations is replaced by the groupoid of changes
of cellular decompositions. We introduce cellular moves which
generate this groupoid and allow to define a renormalization groupoid
flow.

We proceed to test our approach on several models.
Quantum BF theory is the simplest example as it is almost topological
and the renormalization almost trivial. More interesting is
generalized lattice gauge theory for which a qualitative picture of
the renormalization groupoid flow can be given. This is confirmed by
the exact renormalization in dimension two.

A main motivation for our approach are discrete models of quantum
gravity. We investigate both the Reisenberger and the Barrett-Crane
spin foam model in view of their amenability to a renormalization
treatment. In the second case a lack of tunable local parameters
prompts us to introduce a new model.
For the Reisenberger and the new model we discuss
qualitative aspects of the renormalization groupoid flow. In
both cases quantum BF theory is the UV fixed
point.

%% file: intro.tex
\section{Introduction}

Renormalization is an essential tool both in condensed matter and
high energy physics.
It is necessary to make sense of and understand the properties of
physical models. In the first case and to some extent also in the second,
models are often defined as state sums on discretizations of
space-time (spin systems, lattice gauge theory etc.). Renormalization
consists then in understanding and controlling the behaviour
of a model under changes of the discretization.

The discretizations of space (or space-time) employed in these models
are usually hyper-cubic or other types of regular lattices. This is
justified by the fact that the models are defined on a flat metric
background. To describe a change in discretization it is thus
sufficient to specify a scaling factor. 
Renormalization means to tune
the fundamental parameters (coupling constants etc.) of a model in
such a way depending on the
lattice spacing that suitable physical
observables remain (approximately) unchanged. This is expressed
through an action of the group of 
scale transformations (which is referred to as the
\emph{renormalization group} in this context) on the parameter space.
The \emph{renormalization group flows} are the orbits of this action.
A renormalization group \emph{fixed point} is a model (or a
point in parameter space of a model) which remains invariant under
renormalization group (i.e.\ scale) transformations. This usually
implies that it is invariant under all conformal transformations. 

The question we investigate in this paper is what renormalization
means for models defined on
discretizations of space-time without any metric background
structure. Unsurprisingly, the prime example for such models are
non-perturbative models of quantum gravity. Indeed, after introducing
our general approach such models will be the focus of our
investigation.

The first thing to specify is to say what exactly we mean by a
discretization of space-time. This is less trivial than in flat
background models as we can no longer resort to any sort of regular
lattice. Furthermore, we want to respect the global structure of
a space-time manifold in an exact sense and allow the inclusion of
boundaries (although the latter are not explicitly treated in the
present paper). We use the notion of \emph{cellular decomposition},
i.e.\ decomposition as a CW-complex. This includes the notion of 
\emph{simplicial decomposition}, which is employed in many popular
models of quantum gravity. However, cellular decompositions are more
general and appear much better suited to handle the problem of
renormalization as we shall explain.

Practically every discrete model on a compact topological manifold can
be defined using cellular decompositions.
What is more relevant is that many
models can be naturally defined on \emph{arbitrary} cellular
decompositions. It was shown in \cite{Oe:qlgt} that this includes a
generalization of lattice gauge theory and the topological quantum
field theories of Turaev-Viro \cite{TuVi:inv3,BaWe:invplm} and
Crane-Yetter \cite{CrKaYe:inv4}.

The second step is to describe changes of discretization and identify
a suitable analogue of the renormalization group. In contrast to fixed
background models there is no notion of scale or scale transformation
and indeed no notion of global change of discretization at all.
Instead, we consider all possible cellular decompositions of a manifold
and all possible changes between them, i.e.\ any pair of
decompositions. This leads to
a groupoid structure on (the category of) cellular
decompositions. This
\emph{renormalization groupoid} is the analogue
of the renormalization group of flat background models.
Attached to it are notions of \emph{refinement} and
\emph{coarsening} of cellular decompositions. The latter
expresses the idea of integrating out degrees of freedom.

To get some control on the renormalization groupoid we introduce a set
of coarsening moves which we call \emph{cellular moves}. 
There
are $n$ types of such moves in $n$ dimensions.
(One type of move was already introduced in \cite{Oe:qlgt} while for
the case of dimension three the two others were introduced in
\cite{GiOePe:diagtop}.)
In contrast to
the fixed background case changes of discretization can occur not only
locally, but there are even different ways of making a change ``at a
given place''.
We conjecture that any two cellular decompositions
are related by a sequence of these moves and their inverses. 
(This was proven for dimension three in a piecewise linear context in
\cite{GiOePe:diagtop}.) In 
particular, this means that the moves \emph{generate} the groupoid.

To allow for a non-trivial renormalization a model in our context must
have \emph{local} parameters that the renormalization groupoid will
act on. These are parameters that are associated with certain cells in
a cellular decomposition. We define what a \emph{local action} of the
renormalization groupoid means.

Note that our treatment of renormalization takes no prejudice as
to
whether a discrete structure of space-time is really a physical
phenomenon (as in condensed matter physics) or a mathematical artifact
(as in lattice gauge theory).
In both situations the methods developed here should be applicable
as is the case for the renormalization methods for flat background
models.

Let us also mention that there are models which
as a whole do not assume a metric background, but where a sum over
discretizations is performed which individually carry metric
backgrounds. This is notably the case for the ``dynamical
triangulation'' approach to quantum gravity.
In this case conventional renormalization methods are sufficient and
indeed have been successfully applied \cite{ADJ:qgeom}.

After setting up the general framework we proceed to discuss the
renormalization of specific models.
In general, the identification of suitable observables and the actual
renormalization with respect to them is quite a hard
problem. It goes beyond the scope of the present paper where
we only aim at demonstrating the workings of our renormalization
framework in principle. Instead, we contend ourselves with a
renormalization of the partition function. That is, renormalization is
to keep the partition function fixed under changes of cellular
decomposition.

All the models we consider are state sum models and
can be motivated as path integral quantizations of field
theories. The first two models are discrete gauge theories, for which
a fairly extensive discussion of renormalization can be given. 
The simpler model, quantum BF theory, is almost topological, i.e.\
almost a renormalization groupoid fixed point.
Indeed, we perform the exact renormalization which involves a
non-trivial (but rather simple) action of the renormalization groupoid
on a global parameter. This non-trivial action can be considered the
origin of the well known anomaly. Removing the anomaly by inserting an
appropriate factor leads to a true renormalization groupoid fixed
point.
The quantum group generalization of this model is the
Turaev-Viro TQFT (in dimension three) or the Crane-Yetter
TQFT (in dimension four), see \cite{Oe:qlgt}. 

The second model is discrete quantum Yang-Mills theory (or lattice gauge
theory) generalized to cellular decompositions \cite{Oe:qlgt}. This
\emph{cellular gauge theory}
can be viewed as arising from turning a metric background (of
usual Yang-Mills theory) into local parameters of a background-free
model. Indeed, these local parameters are necessary ingredients of a
theory that is not at all topological. 
We discuss general features of the renormalization groupoid flow
including the ultraviolet and infrared fixed points. 
For the two dimensional case we perform an exact renormalization
(suggested by the exact solvability of lattice
gauge theory in two dimensions). It involves a non-trivial
action of the renormalization groupoid on the local parameters and
confirms the qualitative picture of the general case.

The further models we consider are spin foam models of Euclidean
quantum gravity, originally defined on simplicial decompositions.
They also derive from discrete gauge theories and can
be constructed as modifications of quantum BF theory. For these models
we contend ourselves with a discussion of their amenability to a
renormalization treatment in our sense. This implies firstly considering
the models ``as is'', and secondly proposing suitable modifications.

Necessary requirements for renormalization are the presence of local
parameters (as the models are not topological) and their
defineability on arbitrary cellular decompositions.
The first model considered in this context is the
Reisenberger model \cite{Rei:simpgr}. This has a global parameter
which can be easily localized. On the other hand there appears to be
no obvious generalization of the model to cellular decomposition.
Nevertheless, we are able to sketch some properties of the renormalization
groupoid flow.

The second model of quantum gravity we consider is the Barrett-Crane
model \cite{BaCr:relsnet}. While originally defined on simplicial
decompositions only our formulation extends naturally to arbitrary
cellular decompositions (as already suggested by Reisenberger
\cite{Rei:relvert}). On the other hand, local tunable parameters
are completely absent. This prompts us to propose a new model which
interpolates between quantum BF theory and the Barrett-Crane
model. This makes use of a heat kernel operator which in lattice gauge
theory
interpolates between a strong and weak coupling regime. Here the weak
coupling limit corresponds to quantum BF theory (as in lattice gauge theory)
while the strong coupling limit corresponds to the Barrett-Crane
model. Indeed, the renormalization groupoid flow is surprisingly
similar to that of cellular gauge theory including an ultraviolet fixed
point and an ``almost'' infrared fixed point (which is the
Barrett-Crane model).

The formalism we use to express the discussed models is a diagrammatic
one. It is essentially the formalism introduced in \cite{Oe:qlgt} to
represent morphisms in monoidal categories. We employ here a
simplified form adapted to Lie groups and give a self-contained
description of it. This diagrammatics is strongly related to the
connection formulation of discrete gauge models while it also easily
translates into the spin foam formalism. Moreover, in its general form
it includes the supergroup and the quantum group case.
This implies that much of
our treatment of concrete models in this paper generalizes directly to
supergroups and quantum groups as gauge groups (see in particular
Section~4.2 in \cite{Oe:qlgt}).

The first part of the paper presents our proposal of a framework for
renormalization. In Section~\ref{sec:cellular} cellular
decompositions and the cellular moves are introduced while
Section~\ref{sec:renorm} contains the basic notions of renormalization
groupoid and its action. The second part of the paper is devoted
to applications. It starts with
Section~\ref{sec:circdiag} which serves as a brief review of the
diagrammatic language employed in the
following. Section~\ref{sec:mod1} treats the
renormalization of quantum BF theory and of cellular gauge theory.
Section~\ref{sec:mod2} deals with the question of renormalization
for spin foam models of quantum gravity. Finally,
Section~\ref{sec:conclude} presents some conclusions.

%% file: cellular.tex
\section{Cellular decompositions and moves}
\label{sec:cellular}

In this section we give the necessary background on cellular
decompositions and introduce the cellular moves. The former serve as
our definition of ``discretization of a space'', while the latter
serve to formalize and control ``changes of discretization''.

Roughly speaking, a cellular decomposition is a division of a compact
manifold into open balls, called cells. For a manifold of dimension
$n$ there are not only cells of dimension $n$ but also cells of lower
dimension, filling the gaps between the higher dimensional cells, down
to dimension $0$-cells (points).
More precisely, a cellular
decomposition is a presentation of a manifold as a CW-complex. This
can also be formalized as in the following definition.

\begin{dfn}
Let $M$ be a compact manifold of dimension $n$. A \emph{cellular
decomposition} of $M$ is a presentation of $M$ as the disjoint union
of finitely many sets
\[
 M=\bigcup_{k\in\{0,\dots, n\}, i} C^k_{i}
\]
with the following properties: (a) $C^k_i$, called a $k$-\emph{cell}, is
homeomorphic to an open ball of dimension $k$. (An open ball of
dimension 0 is defined to be a point.)
(b) The boundary of each cell is contained in the union of the
cells of lower dimension.
Here, the \emph{boundary} $\partial C$ of a cell $C$ is defined to be
the closure of $C$ in $M$ with $C$ removed.
\end{dfn}

Next, we introduce the concepts of \emph{refinement} and its opposite,
\emph{coarsening}. This is rather intuitive and the definition
straightforward. 

\begin{dfn}
Let $M$ be a compact manifold with cellular decompositions $C$ and
$D$. If each cell in $C$ is equal to some union of cells in $D$ then
$D$ is called a \emph{refinement} of $C$ and $C$ is called a
\emph{coarsening} of $D$.
\end{dfn}

We turn to the \emph{cellular moves}. These are local changes of a
cellular decomposition and they occur in different types.
In dimension
$n$, there are $n$ types of moves together with their inverses.
All
types of moves coarsen a cellular decomposition, while their inverses
refine it. The
$n$-move can be thought of as removing a boundary (an $n-1$ cell)
between two $n$-cells so that they ``fuse'' to one $n$-cell.
The
other moves can all be thought of as removing lower dimensional cells
from the ``interior'' of an $n$-cell.

\begin{dfn}
\label{dfn:moves}
Let $M$ be a manifold of dimension $n$ with cellular decomposition and
$1\le k\le n$.
Let $\sigma$, $\tau$, $\mu$ be respectively a $n$, $k$, $k-1$ cell such
that $\mu$ is contained in the boundary of only two cells: $\sigma$
and $\tau$. The union $\sigma\cup\tau\cup\mu$ is then an open
$n$-ball. Thus we can remove $\sigma$, $\tau$ and $\mu$ from the cellular
decomposition and add the new $n$-cell $\sigma\cup\tau\cup\mu$ instead.
This gives rise to a new cellular decomposition of $M$.
This process is called a \emph{$k$-cell move}, or a move of \emph{type $k$}.
The moves of type $1$ to $n$
are called the
\emph{cellular moves} in dimension $n$.
\end{dfn}

This generalizes the definition of the three moves introduced for
$n=3$ in \cite{GiOePe:diagtop}. (There, the 3-move was called ``3-cell
fusion'', the 2-cell move was called ``2-cell retraction'' and the
1-move was called ``1-cell retraction''.) The $n$-move in any
dimension was already introduced in \cite{Oe:qlgt}.

The crucial point about the moves is that for any two cellular
decompositions of a compact manifold we conjecture that there exists a
sequence of cellular moves that converts one into the other.

\begin{con}
\label{con:cellrel}
Let $M$ be a compact manifold of dimension $n$ with cellular
decompositions $C$ and $D$. Then, $C$ and $D$ are related, up to
cellular homeomorphism, by a
sequence of cellular moves and their inverses in dimension $n$.
\end{con}

While this is still a conjecture, it is true at least in dimension
less than four for piecewise linear manifolds as was shown in
\cite{GiOePe:diagtop}.

Note that many popular models are definied on \emph{simplicial
decompositions} of a manifold. A simplicial decomposition is a special
case of a cellular decomposition where each cell is a
simplex. For simplicial decompositions there is a set of moves, called
the \emph{Pachner moves} which relates any two decomposition of a
given manifold \cite{Pac:equivshell}. Conjecture~\ref{con:cellrel}
above indeed can be considered an analog for cellular
decompositions of Pachner's result.

One advantage of the cellular
moves over the Pachner moves is that they are more
elementary. That a Pachner move can be decomposed into cellular moves
is implied by Conjecture~\ref{con:cellrel}. More importantly,
for BF theory (which is the prototype for all models considered here)
the cellular moves correspond to certain elementary
identities of the partition function (see
Section~\ref{sec:bfren}). Furthermore, it turns out to be 
crucial for renormalization
that the cellular moves are coarsening or (their inverses) refining, 
while most Pachner moves are neither (see the corresponding remark in
Section~\ref{sec:actrgrp}).
On the other hand, it is usually not a big problem to generalize a
model from simplicial to cellular decompositions.

In addition to a cellular decomposition itself it is often convenient
for the formulation of certain models to consider its \emph{dual
complex}. This is obtained by replacing (in dimension $n$) a $k$-cell
by an $n-k$-cell. Usually, we only need the 0, 1 and 2-cells of this
dual complex. 
To distinguish them from the original cells we call them
\emph{vertices}, \emph{edges} and \emph{faces}.
The subcomplex consisting of these is also called the \emph{2-skeleton}
of the dual complex.

%% file: rengrp.tex
\section{The renormalization groupoid}
\label{sec:renorm}

In this section we consider the ensemble of all cellular
decompositions of a manifold. We equip it with a groupoid structure
and explain how this groupoid plays a role in renormalization
analogous to
the group of scale transformations in models with fixed backgrounds.

\subsection{Changes of cellular decomposition}

When renormalizing, we are interested in understanding the behaviour
of a model under a change of discretization.
For fixed-background models a single parameter is
sufficient to describe this, a scale factor. In contrast, there is no
general way to compare arbitrary cellular decompositions without a
background. We thus resort to the most generic way of describing a
change of cellular decomposition, namely by simply specifying 
the initial and 
the final one.

Think of a change of cellular decomposition as an arrow. Arrows can be
composed if the final decomposition of the first one coincides with
the initial decomposition of the second one. Furthermore, there is an
identity arrow from each cellular decomposition to itself. Also, for
each arrow there is an inverse arrow, just because we consider all
possible changes of cellular decomposition, and that includes each
one's inverse. So the arrows (changes of cellular decompositions) form
a groupoid.

\begin{dfn}
Let $M$ be a compact manifold of dimension $n$. Consider the set $C$
of homeomorphism classes of cellular decompositions of $M$. We make
this set into a category as follows. For any two objects $A,B$ we
define exactly one arrow $A\to B$. It is inverse to the arrow $B\to
A$. This category is a groupoid which we call the \emph{cellular
groupoid} of $M$.
\end{dfn}

In terms of renormalization the cellular groupoid plays the role of the
\emph{renormalization groupoid} (replacing the \emph{renormalization
group}).
The cellular moves (and their inverses) appear as particular elements
in the groupoid. What is more, Conjecture~\ref{con:cellrel} implies
that they \emph{generate} this groupoid. Thus, in a sense these moves
can be compared to infinitesimal generators of a transformation group
in a model defined on a background (although they are not infinitesimal
of course, indeed there is no concept of ``infinitesimal'' in our purely
topological setting).

\subsection{Action of the renormalization groupoid}
\label{sec:actrgrp}

In flat background models the action of the renormalization group
is described by the
action of the group of scale transformations $G$ on the space of
parameters (e.g.\ coupling constants) $\Lambda$ of the model.
These parameters are \emph{global} in the sense that they are
associated with the model as such and not with particular places in
space-time.
The action is defined in such a way that it leaves invariant (or
in a suitable sense asymptotically invariant) relevant observables of
the model. The orbits of $G$ in $\Lambda$ are the renormalization
group flows. The renormalization group fixed points are the fixed
points of this action. In the case of scale transformations they
correspond to points in parameter space where the model becomes scale
invariant. This then usually implies conformal invariance.

The analogue of renormalization in our background independent context
is more involved. Again, we should have an action of the
renormalization groupoid on the space of parameters of the model.
However, to consider only global parameters now would be too
restrictive. A local change in discretization changes the model
locally. Thus, to counteract this by renormalization a local
tuning of the model must take place and we need local
parameters. Indeed, even looking at a flat background model there is
often a natural way to localize its parameters. The usual global
nature of the parameters arises then simply due to a degeneracy
introduced by global
space-time symmetries. An example of this is lattice gauge theory (see
Section~\ref{sec:ym}).

Now the action of the cellular groupoid on the space of parameters of
the model should be local. That is, a change in cellular decomposition
should only act on parameters associated with the cells that are
affected by the change. To make this more precise, we consider the
cellular moves and propose the following working definition.

\begin{dfn}
\label{def:locact}
An action of the cellular groupoid on the space of local parameters of
a model is called \emph{local} iff the action of a $k$-cell move
determines the parameters associated to $\sigma'$ and the cells in the
boundary of $\tau$ as a function of the parameters associated to $\sigma$,
$\tau$, $\mu$, and the cells in the boundary of $\tau$, while leaving
other local parameters unchanged. (Terminology of
Definition~\ref{dfn:moves}.)
\end{dfn}

Although we talked about an action of the cellular groupoid so far,
what one usually has is an action of a certain subcatgeory only. The
situation is somewhat analogous to what happens in block spin
transformations. If we coarsen, the number of parameters decreases and
thus information is lost. At the same time the idea is that we
integrate out degrees of freedom. (Note however, that the two effects
are not the same. Parameters are not dynamical degrees of freedom.)
We normally cannot recreate information in a
canonical way. Consequently, it is often not possible to define an
action of certain refinements.

Indeed, the key deficiency of the Pachner moves as a suitable basis of 
a renormalization program is that almost all the moves are not
coarsening in either direction. In other words, at least some moves
would require the creation of information.
In contrast, the cellular moves are all purely coarsening and thus
can only correspond to destruction of information.

As in statistical mechanics
we define the direction of the \emph{renormalization groupoid flow} to
be in the direction of coarsening. 
The action of the cellular moves thus always
points in the direction of the flow, i.e.\ from the ``ultraviolet'' to
the ``infrared''.

%% file: diag.tex
\section{Circuit diagrams}
\label{sec:circdiag}

In this section we recall a few facts about matrix elements of groups
and introduce a convenient diagrammatic language for them.
This diagrammatics then serves as an essential tool
in Sections~\ref{sec:mod1} and \ref{sec:mod2}, where various models
are discussed from the point of view of renormalization.

Related diagrammatic methods for calculations with ``tensors'' go back
a long time.
What is remarkable about the diagrammatics presented here however, is
that it can be generalized to a category theoretic setting
\cite{Oe:qlgt}. This implies for example that much of the treatment in
Sections~\ref{sec:mod1} and \ref{sec:mod2} generalizes
straightforwardly to a setting where gauge groups are replaced by
supergroups or quantum groups.

The diagrammatics introduced here is closely related to the spin
network diagrams traditionally used when working with spin foam
models. We briefly explain their relation at the end of this section.

\subsection{Basics}

Let $G$ be a compact Lie group. We write matrix elements of $G$ as
follows,
\[
 t^V_{i,j}(g)\defeq \langle \phi_i | \rho_V(g) v_j\rangle.
\]
Here, $v_k$ is a basis of the representation space $V$, $\phi_k$ is a
dual basis of the dual space $V^*$ and $\rho_V(g)$ is the
representation matrix for the group element $g$ in the representation
$V$.

Recall that a character $\chi_V$ of a representation $V$ is the
trace $\chi_V(g)=\sum_i t^V_{i,i}(g)$. Furthermore, any class function $f$
of the group, i.e.\ function such that $f(g)=f(hgh^{-1})$ for all $h$,
can be expanded into characters of irreducible representations:
\[
 f(g)=\sum_V f_V \chi_V(g) .
\]

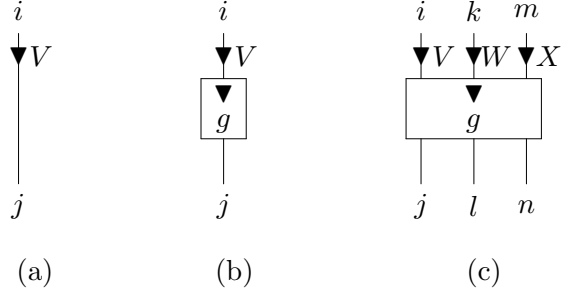
\begin{figure}
\begin{center}
\begin{tabular}{cp{1cm}cp{1cm}c}
\input{figures/fig_matid} &&
\input{figures/fig_mat1} &&
\input{figures/fig_mat3}\\ \\
(a) && (b) && (c)
\end{tabular}
\end{center}
\caption{Diagrams for matrix elements: (a) matrix element at identity
  (a delta function), (b) simple matrix element, (c) product of matrix
  elements.}
\label{fig:diagmatrix}
\end{figure}

Consider now diagrams consisting of lines, called \emph{wires}, and
boxes, called \emph{cables}. Each wire is 
oriented with an arrow and carries the label of a representation of
$G$. Wires can have have free ends and go through cables. Each cable
carries an arrow and is labeled by a group element. The free ends of
wires are labeled by basis indices. Each diagram stands for a matrix
element or for a product of matrix elements as follows:
\begin{itemize}
\item
A wire with free ends is a matrix elements evaluated at the unit
element $e$ of the group. Figure~\ref{fig:diagmatrix}.a represents the
matrix element $t^V_{i,j}(e)=\delta_{i,j}$. The convention is that the
arrow points in the direction from the first index to the second.
\item
A wire going through a cable denotes a matrix element evaluated at the
group element written in the cable. Thus, Figure~\ref{fig:diagmatrix}.b
stands for $t^V_{i,j}(g)$. If the arrows on wire and cable point in
opposite directions the evaluation is instead to be performed at the
inverse of the group element.
\item
Several wires going through a cable correspond to the product of matrix
elements. Figure~\ref{fig:diagmatrix}.c stands for the product
$t^V_{i,j}(g) t^W_{k,l}(g) t^X_{m,n}(g)$.
\item
More complicated diagrams are composed of simpler ones by connecting 
matching ends of wires and contracting the indices.
\end{itemize}
For obvious reasons we call these diagrams \emph{circuit diagrams}.

We also use cables without group labels. Such a cable
stands for the integral over the group of the
respective matrix element(s). Note that in this case also the arrow on
the cable can be unambiguously omitted as the integral is invariant
under inversion. However, the relative orientations of the wires going
through are still important. Only this type of cable will be used in
the following sections. As a side remark, it is only
this type of cable that makes sense in the generalized quantum group
context.

If we consider a diagram that is closed then each wire loop
corresponds to a character evaluated on the product of group elements
labeling the cables traversed by the wire. The simplest closed diagram
consists just of one closed loop of wire. Its value is that
of the dimension of the respective representation.

\begin{figure}
\begin{center}
\input{figures/fig_omega}
\end{center}
\caption{Definition of the formal label $\Omega$ as a sum over
  diagrams. The sum runs over all irreducible representations $V$ and
  is weighted by the dimension.}
\label{fig:omega}
\end{figure}

We also introduce closed
wires with a formal label $\Omega$. This means that one has to perform
a sum over diagrams, with $\Omega$ replaced in each summand by an
irreducible representation. The sum runs over all irreducible
representations and each summand is weighted by the dimension of the
representation, see Figure~\ref{fig:omega}.
This corresponds to a sum over characters which is the delta function:
\[
 \delta(g)=\sum_V \dim V \chi_V(g).
\]
For the closed wires marked with $\Omega$ we can leave out the arrow
as the summation automatically includes dual representations with
equal weight.

Note that as the number of irreducible representations is
infinite, a diagram containing $\Omega$-loops need not represent a
finite quantity. In particular, a single closed $\Omega$-loop
corresponds formally to the infinite quantity
\[
 \kappa\defeq\sum_V (\dim V)^2 .
\]

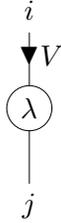
\begin{figure}
\begin{center}
\input{figures/fig_hkdef}
\end{center}
\caption{Heat kernel operator.}
\label{fig:hkdef}
\end{figure}

Another useful type of diagram is a disc  with an arbitrary number of
wires going through. This is the \emph{heat kernel operator}.
For a single wire the disc represents the matrix element
\[
 \langle \phi_i | e^{-\lambda C} v_j\rangle ,
\]
see Figure~\ref{fig:hkdef}.
Here $C$ is the quadratic Casimir
operator. $\lambda$ is a (usually positive real) number which labels
the disc. In the case of several wires the diagram represents the
matrix element on the tensor product representation.

\subsection{Key identities}

In the following sections certain identities of matrix elements and
their integrals play a prominent role. These are conveniently
expressed in the diagrammatic language.

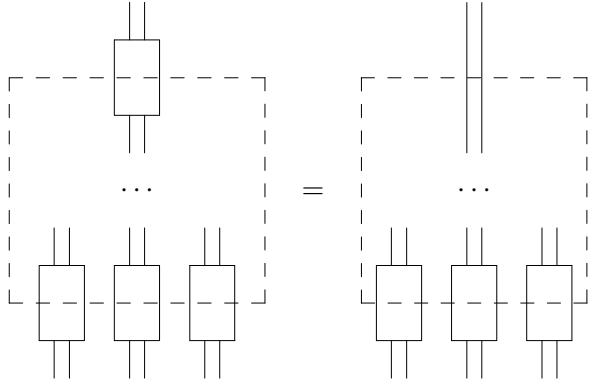
\begin{figure}
\begin{center}
\input{figures/fig_gaugeid}
\end{center}
\caption{Gauge fixing identity.}
\label{fig:gaugefix}
\end{figure}

The first identity of interest is the \emph{gauge fixing identity}.
This is depicted in Figure~\ref{fig:gaugefix}. Consider a circuit
diagram with a closed line inscribed which intersects only cables (the
dashed line in the left hand figure). Then one of the cables can be
removed (exposing the wires) without changing the value of the
diagram. This is true because one integral can be eliminated by
shifting the other integration variables appropriately. In lattice
gauge theory (and in the models we are going to consider in the
following sections) this is
related to gauge fixing \cite{Oe:qlgt}, hence the name of the identity.

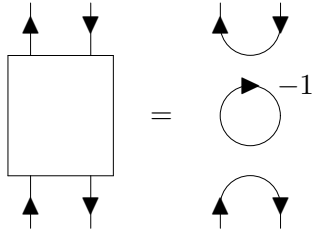
\begin{figure}
\begin{center}
\input{figures/fig_2cableid}
\end{center}
\caption{Tensor product identity for irreducible representations. All
  wires carry the same label.}
\label{fig:tpid}
\end{figure}

Another important identity is the \emph{tensor product identity}. This
can be expressed as
\[
 \int \xd g\, t^V_{i,j}(g) t^V_{k,l}(g^{-1})=\frac{1}{\dim V}
 \delta_{i,l} \delta_{j,k},
\]
where $V$ is an irreducible representation.
The diagrammatic form of the identity is shown in
Figure~\ref{fig:tpid}. Note also that for two inequivalent irreducible
representations the result is zero.

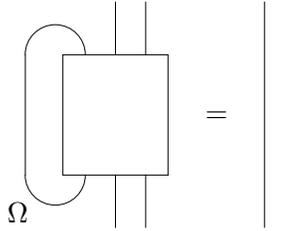
\begin{figure}
\begin{center}
\input{figures/fig_sumid}
\end{center}
\caption{Delta identity. The closed wire carries an $\Omega$-label. Any
 number of wires go through the cable.}
\label{fig:sumid}
\end{figure}

Consider the identity defining the delta function, namely
\[
 \int \xd g\, \delta(g) f(g) = f(e) .
\]
Diagrammatically this is depicted in Figure~\ref{fig:sumid}. We refer
to this for short as the \emph{delta identity}.

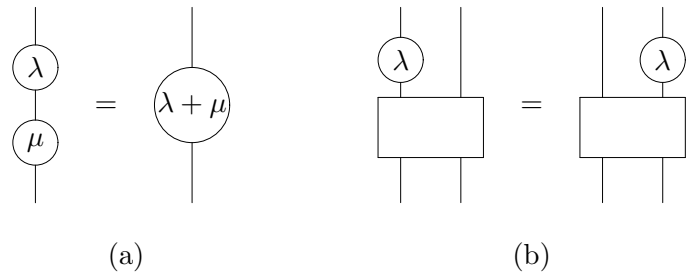
\begin{figure}
\begin{center}
\begin{tabular}{cp{1cm}c}
\input{figures/fig_add} &&
\input{figures/fig_jump}\\ \\
(a) && (b)
\end{tabular}
\end{center}
\caption{Identities involving the heat kernel factor: (a) addition,
(b) jump. The labels on the wires are irrelevant.}
\label{fig:hkid}
\end{figure}

\begin{figure}
\begin{center}
\input{figures/fig_sumhk}
\end{center}
\caption{Modified delta identity with heat kernel factor inserted.
It is obtained by combining the jump identity (Figure~\ref{fig:hkid}.b)
and the usual delta identity (Figure~\ref{fig:sumid}).}
\label{fig:cdelta}
\end{figure}
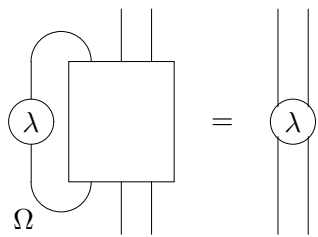

We also exhibit some useful identities involving the heat kernel
operator. The rather obvious fact that $e^{-\lambda C} e^{-\mu
C}=e^{-(\lambda + \mu) C}$ translates into
Figure~\ref{fig:hkid}.a. Also important is the \emph{jump identity},
Figure~\ref{fig:hkid}.b, where each wire can stand for any number of
wires. Furthermore, we can combine the jump and the delta identity,
see Figure~\ref{fig:cdelta}.

What is remarkable is that all these identities (in their diagrammatic
form) continue to hold in a rather general category theoretic context,
including a quantum group setting \cite{Oe:qlgt, GiOePe:diagtop}. (The
heat kernel operator is not mentioned there but one can take in
general an appropriate natural transformation of the identity functor
to obtain its properties.)

\subsection{Relation to spin networks}

The diagrammatic language we use is closely related to the spin
network formalism. Since the latter is the one traditionally used
for spin foam models (as those in Section~\ref{sec:mod2}) we briefly
sketch their relation here.

Our choice to use circuit diagrams
instead of spin networks has two main reasons. The first one is that
for our purposes the formalism of circuit diagrams is considerably
simpler. As already noted in \cite{GiOePe:diagtop}
it avoids complicated identities between arbitrary
$nj$-symbols that would otherwise arise and make a rigorous treatment
rather intractable.
This becomes particularly apparent in the connection
between diagrammatic identities and cellular moves essential for
renormalization in Section~\ref{sec:mod1}. 

The second reason lies in the easy generalization to quantum group
settings. Quantum group spin networks do not share powerful isotopy
properties of circuit diagrams. Consequently they cannot capture
crucial topological information \cite{Oe:qlgt}. This
makes them much less suitable than circuit diagrams for defining and
dealing with models that use quantum groups.

\begin{figure}
\begin{center}
\input{figures/fig_njdec}
\end{center}
\caption{Decomposition of a cable into spin network vertices.}
\label{fig:njdec}
\end{figure}
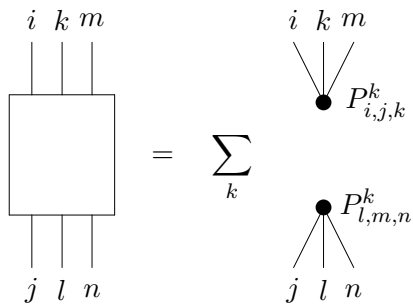

The transition from circuit diagrams to spin networks is essentially
effected by a decomposition of each cable into a pair of spin network
vertices, see Figure~\ref{fig:njdec}. The group
integral represented by the cable can be viewed as projecting the tensor
product of representations labeling the wires onto its trivial
subrepresentation. Now, this trivial subrepresentation can be
decomposed into one-dimensional subspaces.
For example, for a single wire one can write this as
\[
 \int \xd g\, t^V_{i,j}(g)=\sum_k P^k_i P^k_j .
\]
When working with spin networks one consistently chooses such
projectors $P^k_i$ for
every representation (including tensor products).
Diagrammatically one
replaces each end of the cable with a vertex labelled by the
$P^k_i$. The resulting diagram is then a spin network diagram. See
\cite[Section~8.1]{Oe:qlgt} for more details.

%% file: figures/fig_matid.tex
$\rxy{(0,13)*{i},(0,-13)*{j},(0,7)*{\blacktriangledown}
,(0,-10);(0,10)**\dir{-}
,(3,7)*{V}}$

%% file: figures/fig_mat1.tex
$\rxy{(0,13)*{i},(0,-13)*{j},(0,7)*{\blacktriangledown}
,(0,-10);(0,-4)**\dir{-}
,(0,4);(0,10)**\dir{-}
,(-3,-4);(3,-4)**\dir{-};(3,4)**\dir{-}
;(-3,4)**\dir{-};(-3,-4)**\dir{-}
,(0,2)*{\blacktriangledown},(0,-2)*{g}
,(3,7)*{V}
}$

%% file: figures/fig_mat3.tex
$\rxy{(-7,13)*{i},(-7,-13)*{j},(-7,7)*{\blacktriangledown},(-4,7)*{V}
,(0,13)*{k},(0,-13)*{l},(0,7)*{\blacktriangledown},(3,7)*{W}
,(7,13)*{m},(7,-13)*{n},(7,7)*{\blacktriangledown},(10,7)*{X}
,(-7,-10);(-7,-4)**\dir{-}
,(-7,4);(-7,10)**\dir{-}
,(0,-10);(0,-4)**\dir{-}
,(0,4);(0,10)**\dir{-}
,(7,-10);(7,-4)**\dir{-}
,(7,4);(7,10)**\dir{-}
,(-9,-4);(9,-4)**\dir{-};(9,4)**\dir{-}
;(-9,4)**\dir{-};(-9,-4)**\dir{-}
,(0,2)*{\blacktriangledown},(0,-2)*{g}
}$

%% file: figures/fig_omega.tex
\[\rxy{
,(0,-10);(0,10)**\dir{-}
,(2,4)*{\Omega}
}
\quad \defeq \quad \sum_V\quad
\rxy{
,(-12,0)*\cir(4,0){},(-7,-4)*{V}
,(0,-10);(0,10)**\dir{-}
,(2,4)*{V}
}\]

%% file: figures/fig_hkdef.tex
$\rxy{(0,13)*{i},(0,-13)*{j},(0,7)*{\blacktriangledown}
,(0,-10);(0,-3)**\dir{-}
,(0,3);(0,10)**\dir{-}
,(0,0)*\cir(3,0){}*{\lambda}
,(3,7)*{V}
}$

%% file: figures/fig_gaugeid.tex
$\rxy{,(0,0)*{\cdots}
,(-1,25);(-1,20)**\dir{-},(1,25);(1,20)**\dir{-}
,(-1,10);(-1,5)**\dir{-},(1,10);(1,5)**\dir{-}
,(-3,20);(3,20)**\dir{-};(3,10)**\dir{-}
;(-3,10)**\dir{-};(-3,20)**\dir{-}
,(-1,-5);(-1,-10)**\dir{-},(1,-5);(1,-10)**\dir{-}
,(-1,-20);(-1,-25)**\dir{-},(1,-20);(1,-25)**\dir{-}
,(-3,-10);(3,-10)**\dir{-};(3,-20)**\dir{-}
;(-3,-20)**\dir{-};(-3,-10)**\dir{-}
,(-11,-5);(-11,-10)**\dir{-},(-9,-5);(-9,-10)**\dir{-}
,(-11,-20);(-11,-25)**\dir{-},(-9,-20);(-9,-25)**\dir{-}
,(-13,-10);(-7,-10)**\dir{-};(-7,-20)**\dir{-}
;(-13,-20)**\dir{-};(-13,-10)**\dir{-}
,(9,-5);(9,-10)**\dir{-},(11,-5);(11,-10)**\dir{-}
,(9,-20);(9,-25)**\dir{-},(11,-20);(11,-25)**\dir{-}
,(7,-10);(13,-10)**\dir{-};(13,-20)**\dir{-}
;(7,-20)**\dir{-};(7,-10)**\dir{-}
,(-17,15);(17,15)**\dir{--};(17,-15)**\dir{--}
;(-17,-15)**\dir{--};(-17,15)**\dir{--}
} \quad = \quad
\rxy{,(0,0)*{\cdots}
,(-1,25);(-1,5)**\dir{-},(1,25);(1,5)**\dir{-}
,(-1,-5);(-1,-10)**\dir{-},(1,-5);(1,-10)**\dir{-}
,(-1,-20);(-1,-25)**\dir{-},(1,-20);(1,-25)**\dir{-}
,(-3,-10);(3,-10)**\dir{-};(3,-20)**\dir{-}
;(-3,-20)**\dir{-};(-3,-10)**\dir{-}
,(-11,-5);(-11,-10)**\dir{-},(-9,-5);(-9,-10)**\dir{-}
,(-11,-20);(-11,-25)**\dir{-},(-9,-20);(-9,-25)**\dir{-}
,(-13,-10);(-7,-10)**\dir{-};(-7,-20)**\dir{-}
;(-13,-20)**\dir{-};(-13,-10)**\dir{-}
,(9,-5);(9,-10)**\dir{-},(11,-5);(11,-10)**\dir{-}
,(9,-20);(9,-25)**\dir{-},(11,-20);(11,-25)**\dir{-}
,(7,-10);(13,-10)**\dir{-};(13,-20)**\dir{-}
;(7,-20)**\dir{-};(7,-10)**\dir{-}
,(-15,15);(15,15)**\dir{--};(15,-15)**\dir{--}
;(-15,-15)**\dir{--};(-15,15)**\dir{--}
}$

%% file: figures/fig_2cableid.tex
$\rxy{
,(-7,8);(7,8)**\dir{-};(7,-8)**\dir{-}
;(-7,-8)**\dir{-};(-7,8)**\dir{-}
,(-4,12)*{\blacktriangle},(-4,8);(-4,15)**\dir{-}
,(4,12)*{\blacktriangledown},(4,8);(4,15)**\dir{-}
,(-4,-12)*{\blacktriangle},(-4,-8);(-4,-15)**\dir{-}
,(4,-12)*{\blacktriangledown},(4,-8);(4,-15)**\dir{-}
}\quad =\quad
\rxy{
*\cir(4,0){},(0,4)*{\blacktriangleright},(6,4)*{-1}
,(0,12)*\cir(4,0){d_u}
,(0,-12)*\cir(4,0){u_d}
,(-4,12)*{\blacktriangle};(-4,15)**\dir{-}
,(4,12)*{\blacktriangledown};(4,15)**\dir{-}
,(-4,-12)*{\blacktriangle};(-4,-15)**\dir{-}
,(4,-12)*{\blacktriangledown};(4,-15)**\dir{-}
}$

%% file: figures/fig_sumid.tex
$\rxy{
,(-7,8);(7,8)**\dir{-};(7,-8)**\dir{-}
;(-7,-8)**\dir{-};(-7,8)**\dir{-}
,(-8,8)*\cir(4,0){u_d}
,(-8,-8)*\cir(4,0){d_u}
,(-12,-8);(-12,8)**\dir{-}
,(0,8);(0,15)**\dir{-}
,(4,8);(4,15)**\dir{-}
,(0,-8);(0,-15)**\dir{-}
,(4,-8);(4,-15)**\dir{-}
,(-13,-13)*{\Omega}
}\quad =\quad
\rxy{
,(0,-15);(0,15)**\dir{-}
,(4,-15);(4,15)**\dir{-}
}$

%% file: figures/fig_add.tex
$\rxy{,(0,5)*\cir(3,0){}*{\lambda}
,(0,-5)*\cir(3,0){}*{\mu}
,(0,-13);(0,-8)**\dir{-}
,(0,-2);(0,2)**\dir{-}
,(0,8);(0,13)**\dir{-}
}
\quad = \quad
\rxy{,(0,0)*\cir(5,0){}*{\lambda+\mu}
,(0,-13);(0,-5)**\dir{-}
,(0,5);(0,13)**\dir{-}
}$

%% file: figures/fig_jump.tex
$\rxy{,(-4,6)*\cir(3,0){}*{\lambda}
,(-4,-13);(-4,-7)**\dir{-}
,(-4,1);(-4,3)**\dir{-}
,(-4,9);(-4,13)**\dir{-}
,(4,-13);(4,-7)**\dir{-}
,(4,1);(4,13)**\dir{-}
,(-7,-7);(7,-7)**\dir{-};(7,1)**\dir{-}
;(-7,1)**\dir{-};(-7,-7)**\dir{-}
}
\quad = \quad
\rxy{,(4,6)*\cir(3,0){}*{\lambda}
,(-4,-13);(-4,-7)**\dir{-}
,(-4,1);(-4,13)**\dir{-}
,(4,-13);(4,-7)**\dir{-}
,(4,1);(4,3)**\dir{-}
,(4,9);(4,13)**\dir{-}
,(-7,-7);(7,-7)**\dir{-};(7,1)**\dir{-}
;(-7,1)**\dir{-};(-7,-7)**\dir{-}
}$

%% file: figures/fig_sumhk.tex
$\rxy{
,(-7,8);(7,8)**\dir{-};(7,-8)**\dir{-}
;(-7,-8)**\dir{-};(-7,8)**\dir{-}
,(-8,8)*\cir(4,0){u_d}
,(-8,-8)*\cir(4,0){d_u}
,(-12,-8);(-12,-3)**\dir{-}
,(-12,3);(-12,8)**\dir{-}
,(0,8);(0,15)**\dir{-}
,(4,8);(4,15)**\dir{-}
,(0,-8);(0,-15)**\dir{-}
,(4,-8);(4,-15)**\dir{-}
,(-12,0)*\cir(3,0){}*{\lambda}
,(-13,-13)*{\Omega}
}\quad =\quad
\rxy{
,(0,-15);(0,-2)**\dir{-}
,(4,-15);(4,-2)**\dir{-}
,(0,2);(0,15)**\dir{-}
,(4,2);(4,15)**\dir{-}
,(2,0)*\cir(3,0){}*{\lambda}
}$

%% file: figures/fig_njdec.tex
\[\rxy{
,(-7,8);(7,8)**\dir{-};(7,-8)**\dir{-}
;(-7,-8)**\dir{-};(-7,8)**\dir{-}
,(-4,8);(-4,15)**\dir{-}
,(4,8);(4,15)**\dir{-}
,(0,8);(0,15)**\dir{-}
,(0,-8);(0,-15)**\dir{-}
,(-4,-8);(-4,-15)**\dir{-}
,(4,-8);(4,-15)**\dir{-}
,(-4,18)*{i},(-4,-18)*{j},(0,18)*{k},(0,-18)*{l}
,(4,18)*{m},(4,-18)*{n}
}\quad =\quad \sum_k\quad
\rxy{
,(0,7)*\frm<3pt>{*};(0,-7)*\frm<3pt>{*} 
,(7,7)*{P^k_{i,j,k}},(7,-7)*{P^k_{l,m,n}}
,(0,7);(-4,15)**\dir{-}
,(0,7);(0,15)**\dir{-}
,(0,7);(4,15)**\dir{-}
,(0,-7);(-4,-15)**\dir{-}
,(0,-7);(0,-15)**\dir{-}
,(0,-7);(4,-15)**\dir{-}
,(-4,18)*{i},(-4,-18)*{j},(0,18)*{k},(0,-18)*{l}
,(4,18)*{m},(4,-18)*{n}
}\]

%% file: models1.tex
\section{Discrete gauge theory models}
\label{sec:mod1}

The background-free models we study in this paper are all
(essentially) discrete quantum gauge theories.
Prototypical and well suited to test
our approach to renormalization are quantum BF theory and a
generalization of discrete quantum Yang-Mills theory (lattice gauge
theory). Both are treated in the present section.

Quantum BF theory is essentially topological and we show how this
comes out in our approach. We then show how the anomaly that
prevents true topological invariance arises from a non-trivial action
of the renormalization groupoid on a global parameter. From this the
known change of normalization can be derived that leads
to a topological theory, i.e.\ to a renormalization groupoid fixed point.

Discrete quantum Yang-Mills theory
is more interesting from the point of view of
renormalization.
Removing the background, the information about the metric condenses
into local parameters.
One obtains a generalization of lattice gauge theory which might be
called \emph{cellular gauge theory}.
It contains BF theory in its parameter space as a
weak coupling limit.
We discuss general features of the renormalization groupoid flow.
While an open problem in higher dimensions, an exact renormalization
can be carried out in dimension two. We show how this involves a
non-trivial action of the renormalization groupoid on the local
parameters.

We restrict ourselves in the following to considering no observables
but partition functions only. That is, renormalization is to be
understood purely as renormalization with respect to keeping the
partition function fixed.

\subsection{Quantum BF theory}
\label{sec:bf}

\subsubsection{Discretization and quantization}

While really being interested in the discrete model (\ref{eq:bfpart})
we start by briefly recalling its motivation as the quantization of
continuum BF
theory. For more details on this subject we recommend
\cite{Bae:introsfoam}.

Let $M$ be a compact manifold of dimension $n$, $G$ a compact Lie
group and $P$ a principal $G$-bundle over $M$. We consider a
connection $A$ on $P$ and an $n-2$ form $B$ with values in the vector
bundle associated to $P$ via the adjoint action. Define the action
\begin{equation}
 S=\int_M \tr(B\wedge F) ,
\label{eq:bfact}
\end{equation}
where $F$ is the curvature 2-form of $A$ and $\tr$ is the trace in the
fundamental representation.

We perform path integral quantization to obtain the partition
function 
\[
 \pf=\int \xD A\, \xD B\, e^{\im S} .
\]
Formally integrating out the $B$-field leads to
\[
 \pf=\int \xD A\, \delta(F) ,
\]
i.e.\ we obtain the integral over all connections with vanishing
curvature.

To make sense of this expression we discretize $M$ and proceed as in
lattice gauge theory, i.e.\ by assigning parallel transports to edges,
curvatures (holonomies) to faces etc. However, not having any fixed
background there is no canonical or ``regular'' way of choosing a
discretization.
We need to consider arbitrary discretizations. That is, in the
language introduced above, cellular decomposition of $M$. Let $\cdec$
be such a cellular decomposition and $\dskel$ the 2-skeleton $\dskel$
of the dual complex (see the end of Section~\ref{sec:cellular}). As in
lattice gauge theory we express
the connection $A$ through its parallel transports along edges $e$ of
$\dskel$. The curvature is then measured on each face $f$ through the
holonomy around the face, i.e.\ the product of the parallel transports
around the face. The measure over connections becomes the product of
Haar measures per edge. That is, we obtain the discretized partition
function
\begin{equation}
\pfx{BF}=\int\prod_e \xd g_e \prod_f \delta(g_1 \cdots g_k) .
\label{eq:bfpart}
\end{equation}
Here $\delta$ is the delta function on the group and $g_1$, \dots,
$g_k$ are the group elements associated with the edges that bound the
face $f$. Note that in writing this expression one chooses an
orientation for each edge, an orientation for each face and a starting
vertex in the boundary of each face. However, all these choices are
irrelevant as the delta function is invariant under conjugation and
the Haar measure is invariant under inversion.

We proceed to express the partition function (\ref{eq:bfpart}) in terms
of circuit diagrams \cite{Oe:qlgt,GiOePe:diagtop}: For each edge we
obtain an unmarked cable
representing the associated group integral. For each face we obtain a
wire going around the face through all the cables on the
bounding edges. The wires represent the delta functions and are marked
$\Omega$.

\begin{figure}
\begin{center}
\includegraphics{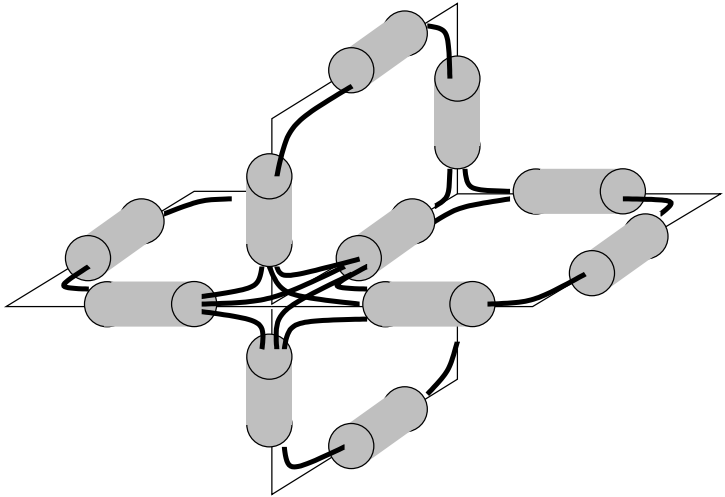}
\end{center}
\caption{Piece of embedded circuit diagram for quantum BF theory. The
  picture shows a three dimensional example. Labels and arrows on
  wires are omitted.}
\label{fig:circdiag}
\end{figure}

We can think of the obtained circuit diagram as embedded into the
manifold. A piece of such an embedded diagram in three dimensions is
shown in Figure~\ref{fig:circdiag}. It is indeed this representation of 
the partition function as a diagram embedded into the manifold which
considerably simplifies the discussion of renormalization.

Explicitly expanding the delta functions in the partition functions in
characters we obtain
\begin{equation}
 \pfx{BF}=\sum_{V_f} (\prod_f \dim_{V_f}) \int\prod_e \xd g_e \prod_f
 \chi_{V_f}(g_1 \cdots g_k) .
\label{eq:bfexp}
\end{equation}
Here the sum runs over all assignments $V_f$ of irreducible
representations to faces $f$. This corresponds diagrammatically to the
expansion in Figure~\ref{fig:omega}.
One might consider this a more proper version of discrete BF theory
than (\ref{eq:bfpart}) as it contains explicit representation valued
degrees of freedom which can be considered the discrete version of the
$B$-field.

Reformulated in the spin foam formalism the
partition function (\ref{eq:bfexp}) yields in dimension three the
Ponzano-Regge model
\cite{PoRe:limracah} and in dimension four the
Ooguri model \cite{Oog:toplat}.

\subsubsection{Exact renormalization}
\label{sec:bfren}

The proof of topological invariance of (anomaly corrected) quantum BF
theory in dimension three using the present formalism of circuit
diagrams and certain cellular moves was already exhibited in
\cite{GiOePe:diagtop}. We extend this here to a full renormalization
treatment and perform the (surprisingly simple) generalization to
arbitrary dimensions.

The first step in understanding the renormalization of quantum BF
theory is to investigate the change of the partition function $\pfx{BF}$
under change of cellular decomposition. Thanks to
Conjecture~\ref{con:cellrel} it is sufficient to regard the change of
$\pfx{BF}$ under the cellular moves.

\begin{table}
\begin{center}
\begin{tabular}{|lcc|}
\hline
move & initial configuration & final configuration \\
\hline
&&\\
$n$&
\includegraphics[scale=0.5]{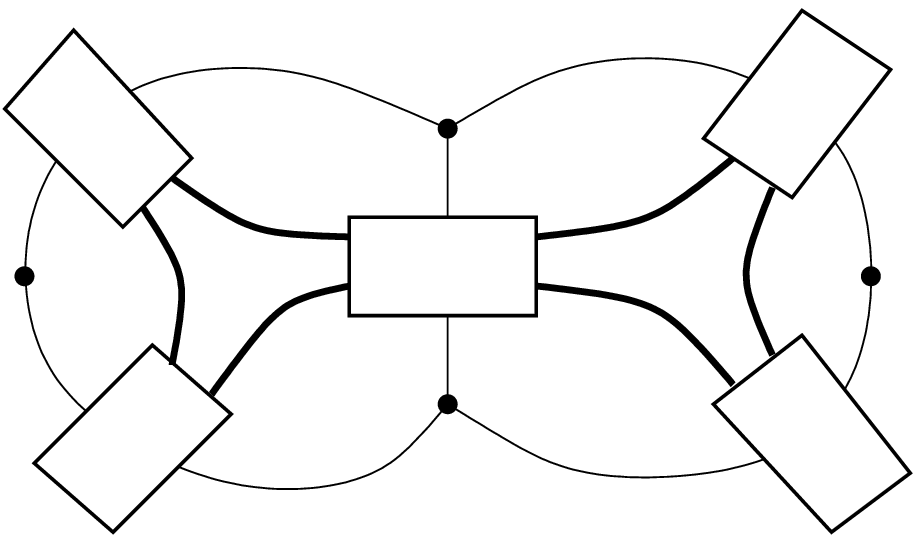} &
\includegraphics[scale=0.5]{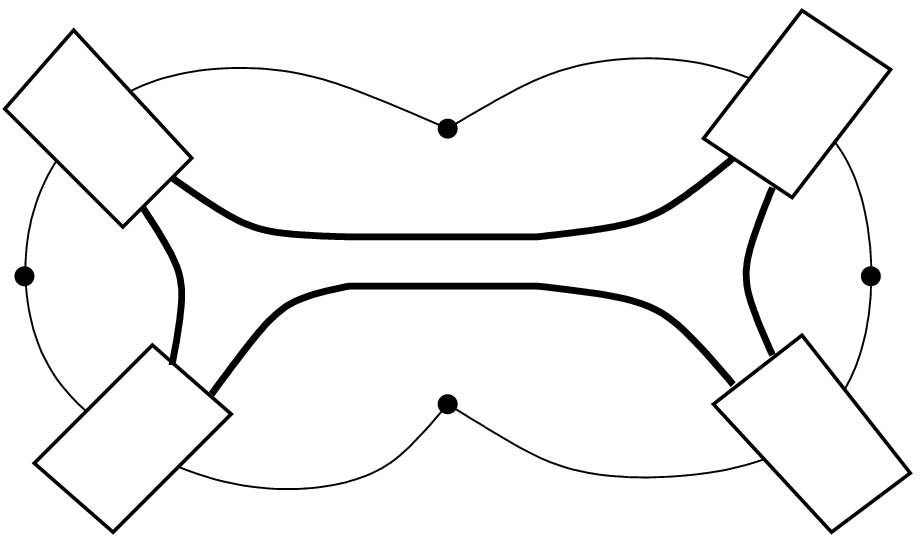} \\
&&\\
\hline
&&\\
$n-1$&
\includegraphics[scale=0.5]{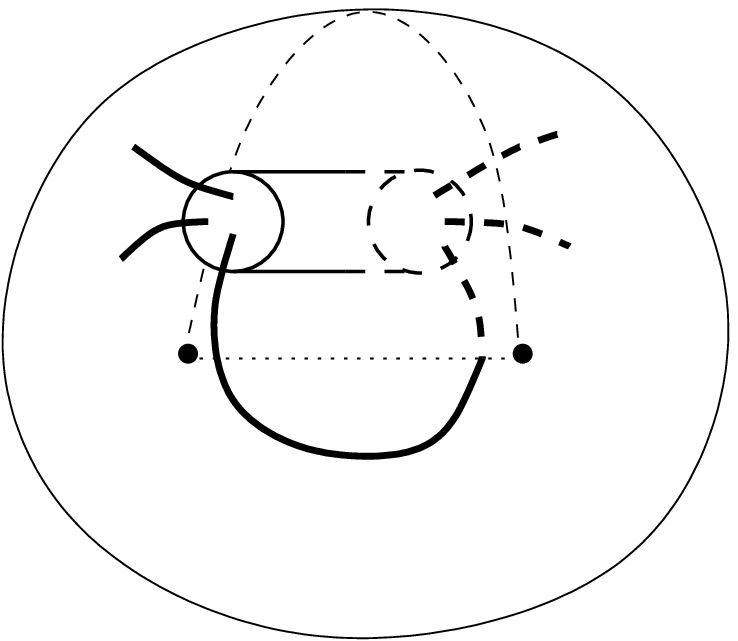} &
\includegraphics[scale=0.5]{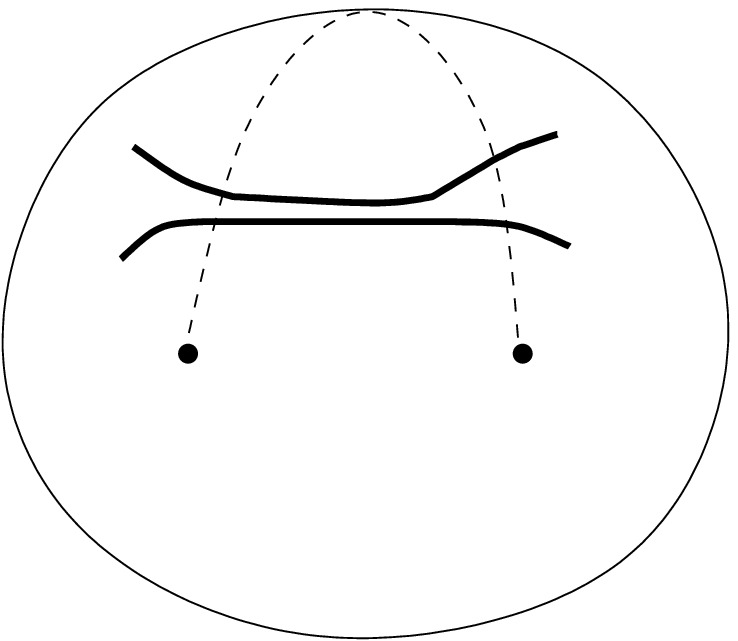} \\
&&\\
\hline
&&\\
$n-2$&
\includegraphics[scale=0.5]{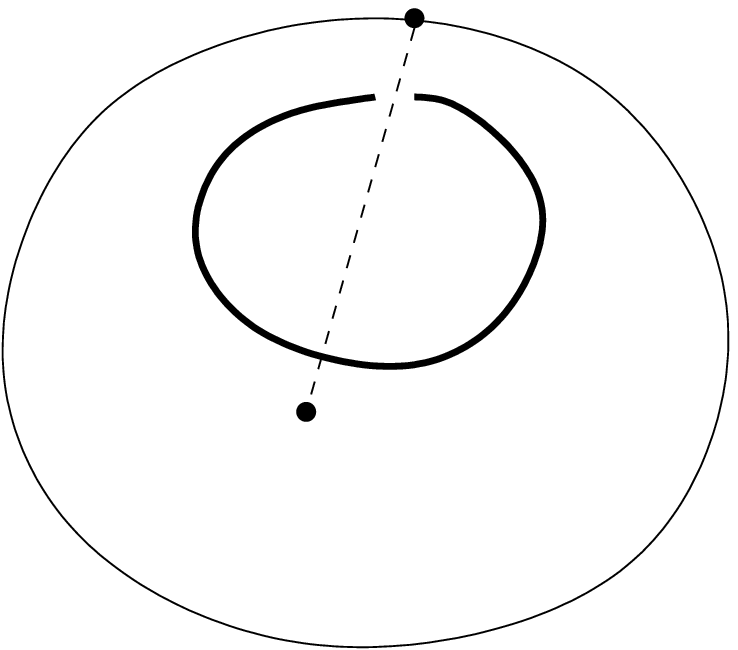} &
\includegraphics[scale=0.5]{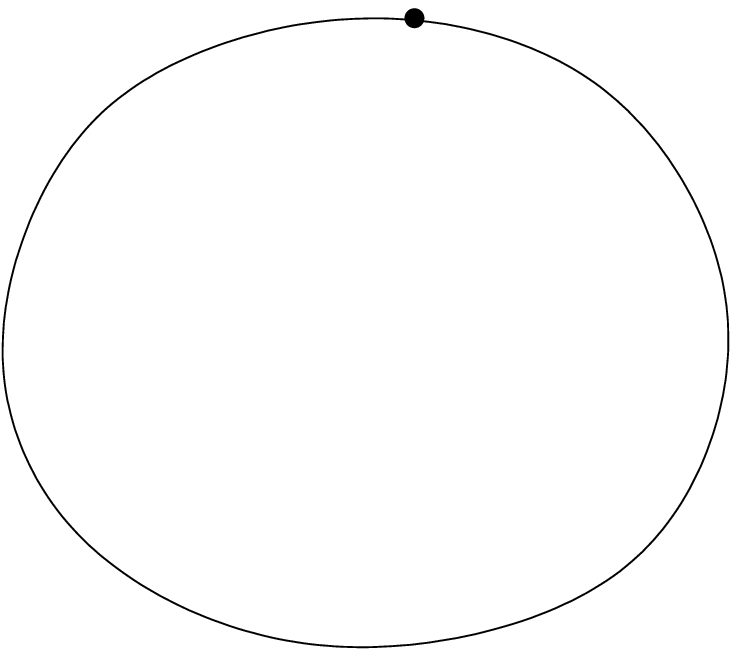} \\
&&\\
\hline
\end{tabular}
\end{center}
\caption{The effect of the cellular moves on the circuit diagrams. The
relevant cells as well as the embedded cables (boxes/cylinders) and
wires (thicker lines) are shown.}
\label{tab:bfmoves}
\end{table}

As $\pfx{BF}$ is encoded in the
embedded circuit diagram constructed above we need to compare the
circuit diagrams before and after the application of the respective
move. This is shown in Table~\ref{tab:bfmoves}. Only the parts of the
circuit diagrams affected by the moves are shown. Furthermore, only
the moves of type $n$, $n-1$ and $n-2$ are indicated. The other moves do
not affect cells of dimension $\ge n-2$ and thus leave the circuit
diagram unchanged.

So, how does the value of $\pfx{BF}$ change? Comparing
Table~\ref{tab:bfmoves} with Figures~\ref{fig:gaugefix} and
\ref{fig:sumid} we find that both the $n$ as well as the $n-1$ move
leave $\pfx{BF}$ invariant. This is due respectively to the gauge fixing
and the delta identity. By contrast, the $n-2$ move removes a
factor corresponding to a closed $\Omega$-loop, i.e.\ a factor of
$\kappa$.\footnote{We shall not worry about the fact that this
factor is divergent. Although we renormalize here with respect to the
partition function, what is finally relevant are physical observables
which should be finite.}

Now we shall consider the renormalization of $\pfx{BF}$, i.e.\ we wish to
construct an action of the cellular groupoid on parameters of the
model such that $\pfx{BF}$ becomes invariant. The model as defined by
(\ref{eq:bfpart}) has no parameters. We need to modify it by
introducing parameters. However, as we have seen we only have to
compensate for factors of $\kappa$ which carry no local information of
the model. It is sufficient to introduce a global integer
parameter $p$ which counts these factors. We define the partition
function of the modified model simply as
\begin{equation}
\pfmx{BF}\defeq \kappa^p \pfx{BF}.
\label{eq:bffixed}
\end{equation}

It remains to define the actions of the moves. We let the
moves of type $n$, $n-1$, $n-3$, $n-4$, \dots act trivially while we let the
move of type $n-2$ act by sending $p\mapsto p+1$ and its inverse by
sending $p\mapsto p-1$. This precisely cancels the factor $\kappa$
appearing in the circuit diagram for the $n-2$-move. We obtain an
exact renormalization.

The renormalized model has a free global parameter $p$. To specify it,
one has to specify its value in a given cellular decomposition. This
determines it, by the action of the renormalization groupoid, for all
cellular decompositions.
From the point of view of the classical continuum model the
discretization allows an exact quantization (i.e.\ not depending on
it) up to the \emph{anomaly} manifest in $p$.

Alternatively, we can fix the extra parameter $p$ by making it explicitly
depend on the numbers $c_k$ of cells of given dimension $k$ (as done
to obtain topological invariants
\cite{TuVi:inv3,BaWe:invplm,CrKaYe:inv4}). For
example we can set\footnote{Other
choices are obtained for example by adding integer multiples of the
Euler characteristic $\chi\defeq c_n- c_{n-1}+c_{n-2}-\dots$ of
$M$. In particular, the choice in \cite{TuVi:inv3,BaWe:invplm} is
$p+\chi=-c_0$ while the choice in
\cite{CrKaYe:inv4} is $p+\chi=-c_1+c_0$.}
\[
 p\defeq -c_n+c_{n-1}-c_{n-2} .
\]
It is easy
to see that $p$ defined in this way is changed only by moves of type $n-2$,
and exactly in the right way. Indeed we can make this part of the
definition of the model instead of its renormalization. The action of
the renormalization groupoid is then trivial.
The model is then
independent of the cellular decomposition, it is a
\emph{renormalization groupoid fixed point}.\footnote{We suppose here
  as everywhere the validity of
  Conjecture~\ref{con:cellrel}. However, something weaker is
  sufficient for our reasoning to hold. Compare the respective remarks
  in Section~\ref{sec:conclude}.}
When referring to quantum BF theory in the following we will usually mean
this version (\ref{eq:bffixed}) where the anomaly has been fixed in
some suitable way.

\subsection{Generalized Yang-Mills theory}
\label{sec:ym}

To talk about renormalization might seem somewhat artificial in
the context of quantum BF theory. Indeed we saw that there is only
one global parameter involved which can be easily absorbed into a
redefinition of the partition function. As a less trivial example we
discuss Yang Mills theory in this section.

It might seem surprising that we apparently wish to consider a
theory that is defined on a metric background. However, we really
consider a generalization of Yang-Mills theory that naturally
arises in a discretized setting \cite{Oe:qlgt}. This provides a nice
example of how a background structure can be turned into local
parameters. These local parameters are then precisely what
the renormalization groupoid acts on.

\subsubsection{Discretization and quantization}

We start again from a continuum formulation.
Let $M$ be a compact
manifold of dimension $n$, $G$ a compact Lie group and $P$ a principal
$G$-bundle over $M$. Imagine we have some theory, determined through
an action $S$ which depends on a connection $A$ on $P$.
In order to define a quantum theory via a path integral\footnote{Note
that we choose a Euclidean path integral. This is in accordance with
standard practice in lattice gauge theory. We are not interested in
physical implications of such choices here.}
\[
 \pf=\int \xD A\, e^{-S} ,
\]
we
discretize the manifold $M$ via a cellular decomposition $\mathcal{C}$
of $M$.

We proceed as above, i.e.\ we discretize the connection by associating
parallel transports to edges $e$ and holonomies to faces $f$ of the
2-skeleton of the dual complex. The most general local\footnote{Local
means here that interactions take place only within each face. In
lattice gauge theory terminology this is called ``ultra-local''.}
gauge
invariant action is then given by
\[
 S=\sum_f \sigma_f (g_1\cdots g_k) ,
\]
where $g_1,\dots, g_k$ are the group elements associated to the edges
that bound the face $f$. $\sigma_f$ is a choice of class
function for each face $f$.
This yields the partition function
\[
\pfx{CGT}=\int\prod_e \xd g_e \prod_f \rho_f(g_1 \cdots g_k),
\]
where we set $\rho_f\defeq e^{-\sigma_f}$.

Indeed we see immediately that it is a generalization of the quantum
BF theory partition function (\ref{eq:bfpart}). Decomposing $\rho_f$
into characters 
\[
 e^{-\sigma_f}=\sum_V \alpha_{V,f} \chi_V ,
\]
yields
\begin{equation}
 \pfx{CGT}=\sum_{V_f}(\prod_f \alpha_{V_f,f}) \int \prod_e \xd g_e
 \prod_f \chi_{V_f}(g_1\cdots g_k) .
\label{eq:gaugepart}
\end{equation}
Here the sum over irreducible representations for each face is taken
to the front as a sum over assignments of an irreducible representation
$V_f$ to each face $f$ (compare (\ref{eq:bfexp})).
As (\ref{eq:gaugepart}) is a generalization of lattice gauge theory to
cellular decompositions we call it \emph{cellular gauge theory} in the
following.

From this expression we can read off
the diagrammatic representation of $\pfx{CGT}$ \cite{Oe:qlgt}. The circuit
diagram is exactly the same embedded graph as for BF theory
(Figure~\ref{fig:circdiag}). Only now the summation over irreducible
representations for each closed wire (corresponding to each face)
cannot be hidden in $\Omega$ labels for the wires.
Instead the wires must carry explicit labels and the sum over diagrams
is performed with the weight factor $(\prod_f \alpha_{V_f,f})$.

Note that the partition function (\ref{eq:gaugepart}) has locally
infinitely many parameters. For each face $f$ there is a choice of
parameter $\alpha_{V,f}$ for each irreducible representation $V$ (of
which there are infinitely many). In the following we restrict
ourselves to a more manageable situation. We set
\begin{equation}
 \alpha_{V,f}\defeq (\dim V)\, e^{-\lambda_f C_V},
\label{eq:sethk}
\end{equation}
where $C_V$ is the value of the quadratic Casimir operator on $V$ and
$\lambda_f$ is a positive real parameter depending on the face.

We now turn to the relation with Yang-Mills theory. Assume that $M$ is
equipped with a flat metric and that $\mathcal{C}$ is a cellular
decomposition of $M$ into hypercubes of equal side length $a$. We set
all $\lambda_f$ equal to $\lambda \defeq a^{n-4} \gamma^2$. Then it
can be shown 
that in the limit $a\to 0$ the action defined in this way
approximates
the continuum action of Yang-Mills theory (up to a constant)
\[
 S_{\text{YM}} = -\frac{1}{\gamma^2}\int_M \tr(F\wedge \star F) ,
\]
with coupling constant $\gamma$.
Indeed, this is at the foundation of lattice gauge theory (see e.g.\
\cite{Cre:lgt}) and the discrete action we arrived at is nothing but
the \emph{heat kernel action} of lattice gauge theory.
Note that that $\lambda$ plays the role of (the square of) a
coupling constant.

\begin{figure}
\begin{center}
\input{figures/fig_hklim}
\end{center}
\caption{The two limits of the heat kernel operator.}
\label{fig:limhk}
\end{figure}
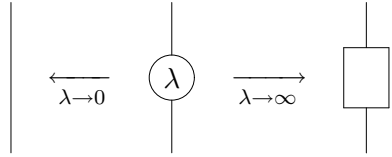

Observe that the heat kernel operator $e^{-\lambda C}$ has two
interesting limits,
$\lambda\to 0$ and $\lambda\to \infty$. In the first case it just
becomes the identity. In view of the above this implies that quantum
BF theory can be considered the weak coupling limit ($\lambda=0$) of
lattice gauge theory, since we then recover (\ref{eq:bfexp}) from
(\ref{eq:gaugepart}). The opposite regime ($\lambda\to\infty$) is that
of strong coupling of lattice gauge theory. The low dimensional
representations dominate more and more in the partition function. At
the extreme point $\lambda=\infty$, only the trivial representation
contributes, onto which the heat kernel operator becomes a projector.
Diagrammatically the limits can be expressed as in
Figure~\ref{fig:limhk}.

Let us go back to the case of interest, that of a topological manifold
$M$ with arbitrary cellular decomposition $\mathcal{C}$.
Note that the above relation also suggests to view $\lambda_f$ as a
remnant of a local metric. 
As we shall see below this
statement can be made precise in the case of dimension two.

Diagrammatically, the choice (\ref{eq:sethk}) means that we can put
the heat kernel factor per face $f$ into the circuit diagram as a disc with
label $\lambda_f$. The summation over irreducible representations with
the remaining factor of
dimension can then be indicated by $\Omega$-labels for all the wires.

\subsubsection{Renormalization in general}
\label{sec:rencgt}

We start by investigating the change in $\pfx{CGT}$ under the moves. Using
the diagrammatic representation this step is almost identical to what
we did for quantum BF theory. Indeed, again the relevant parts of the
diagrams
are as shown in Table~\ref{tab:bfmoves}. The difference is that we now
sum over irreducible representations for each wire with a weight that
is not the dimension, as can be indicated by an extra disc diagram
per wire.

This makes no difference for the $n$-move, as
the gauge fixing identity (Figure~\ref{fig:gaugefix}) does not depend
on attached labels. 
For the $n-1$-move, however, the delta identity
(Figure~\ref{fig:sumid}) that ensured invariance in the BF case can no
longer be applied. The weight in the summation over representations
for the closed wire is no longer the dimension. For the $n-2$-move
there is again just a mismatch of a factor, which is now
\[
 \kappa_f\defeq \sum_V (\dim V) \alpha_{V_f,f}=\sum_V (\dim V)^2
 e^{-\lambda_f C_V}. 
\]
In summary, the most crucial difference to quantum BF theory is the
breaking of invariance under the $n-1$-move.

Let us turn to the problem of renormalization. As before, the treatment of
the $n-2$-move is relatively easy. Redefining the partition function
to be
\[
 \pfmx{CGT}\defeq (\prod_f \kappa_f^{-1}) \pfx{CGT} ,
\]
fixes the $n-2$-move, i.e.\ makes $\pfmx{CGT}$ invariant under it.
On the other hand, this makes the non-invariance under the $n-1$-move
worse in the sense that it would now be broken even in the limit
$\lambda_f=0$. To remedy this and obtain the renormalization groupoid
fixed point of quantum BF theory in this limit one could include a
factor of $\kappa^{-c_n+c_{n-1}}$ (compare Section~\ref{sec:bfren}).

The renormalization of the $n-1$-move poses much deeper
problems. Superficially it seems that instead of applying the delta
identity, we have to apply the modified delta identity
(Figure~\ref{fig:cdelta}). Indeed we can do this and perhaps this is
really a first step towards tackling the general problem. However, the
circuit diagram we arrive at is not of the same kind as the original
circuit diagram: The shifted heat kernel diagram extends over several
wires.

Although no general solution to the problem is known, qualitative
features of the renormalization groupoid flow are easily
described. Firstly, the parameter space of cellular gauge theory
contains two fixed points (we
assume the anomaly has been fixed as described above). The first one, at
$\lambda=0$ (weak coupling limit) is quantum BF theory while the
second one at $\lambda=\infty$ (strong coupling limit) is a trivial
theory. (In the latter all
representations are trivial and invariance under the moves is
trivially satisfied.)

Secondly, we can say something about the
direction of the renormalization groupoid flow. Consider a region of
the manifold with some given parameter values $\lambda_f$. Roughly
speaking, we can perform a refinement without changing the partition
function by assigning to the newly created parameters the topological
value $\lambda_f=0$. Now the partition function in the refined region
should be left unchanged if we average the parameters $\lambda_f$ in
this region in some suitable way. That is, the average value of the
parameters (of which there are now more) has decreased compared to the
unrefined situation. In summary, if we
coarsen, the parameters $\lambda_f$ must generally increase to keep
the partition function fixed. The renormalization groupoid flow goes
in the direction
from lower values of $\lambda_f$ in the ultraviolet to higher
values in the infrared.

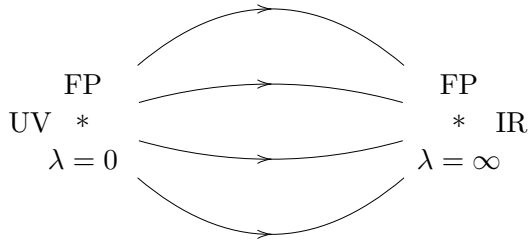
\begin{figure}
\begin{center}
\input{figures/fig_rgflow}
\end{center}
\caption{Qualitative picture of the renormalization groupoid flow for
  cellular gauge theory (generalized discrete quantum Yang-Mills
  theory). The two fixed-points are quantum BF theory in the weak
  coupling limit
  ($\lambda=0$) and the trivial theory in the strong coupling limit
  ($\lambda=\infty$). The general direction of the flow is from the
  first (UV) fixed-point to the second (IR).}
\label{fig:rgflow}
\end{figure}

This general behaviour applies in particular close to the
renormalization groupoid fixed points. (Consider for example a
situation where only a few $\lambda_f$ are non-zero or non-infinite.)
Thus, quantum BF theory is a (repulsive) UV fixed point while the
trivial theory is an (attractive) IR fixed point.\footnote{Note a
  subtle difference to the conventional situation with continuous
  scale transformations. There, one would get arbitrarily close to the
  IR fixed point by further and further coarsening (increasing the
  scale). In our case the manifold is compact and there are
  maximally coarse decompositions. Consequently, for any given
  starting point (which is not the IR fixed point) the renormalization
  groupoid flow stops at a finite distance from the IR fixed
  point.}
This is
schematically illustrated in Figure~\ref{fig:rgflow}.

\subsection{Exact renormalization in dimension two}

In the case of dimension two an exact renormalization can be
performed which confirms the qualitative picture.
As lattice gauge theory is solvable then \cite{Mig:reclgt} this is not
at all surprising.
That things can be made well in this case is rather obvious in our
approach. As each cable has only two wires (because each edge bounds
only two faces), the shifted heat kernel factor again sits on a single
wire.

As there is no $n-2$-move in dimension $n=2$ we consider the original
partition function $\pfx{CGT}$ (and not $\pfmx{CGT}$). Now, after
applying the modified delta identity, we can combine the shifted heat
kernel factor with the one that was attached to the wire before via
addition (Figure~\ref{fig:hkid}.a). We simply get the
heat kernel factor for the sum of the parameters.

The solution of the renormalization problem in dimension two is thus as
follows. The 2-move acts trivially. The 1-move acts by sending the
pair of parameters $(\lambda_{f_1},\lambda_{f_2})$ to
$\lambda'_{f_2}=\lambda_{f_1}+\lambda_{f_2}$. Here, $f_1$ denotes the
0-cell (or face) that is removed, $f_2$ denotes the other 0-cell (or
face) with which $f_1$ was connected through a 1-cell (edge). The new
parameter value $\lambda'_{f_2}$ replaces the old one for $f_2$. The
so defined action leaves $\pfx{CGT}$ invariant and is local in the
sense of Definition~\ref{def:locact}.

Note that the action is not defined for the whole cellular
groupoid, but only in the direction of coarsening. (An action of the
inverse 1-move is not defined.) This is due to the
impossibility of unambiguously recreating local parameters
and goes hand in hand with
the idea of integrating out degrees of freedom
(see the discussion
at the end of Section~\ref{sec:actrgrp}).

Coming back to the interpretation of cellular gauge theory as
Yang-Mills theory we can give the local parameters indeed a simple
meaning in terms of metric information. We can think of $\lambda_f$ as
(proportional to) the area of the face $f$. The 1-move might then be
thought of as the merging of two faces, thus giving the new face the
sum of the areas.

It is sometimes said that 2-dimensional Yang-Mills theory is
topological. This is not true in our terminology,
i.e.\ it is not a renormalization groupoid fixed
point. However, its renormalization is exact and rather simple. Of
course, it is even simpler in the context of lattice gauge theory
where all local parameters are identical and thus only one parameter
changes under change of scale.

As a side remark, we observe that it is very easy to solve the two
dimensional theory using the diagrammatics \cite{Oe:qlgt}. Just apply
the tensor product identity (Figure~\ref{fig:tpid}) to all
cables. Only closed loops of wire without any cables are
left. Counting the loops and taking care of the heat kernel factors
one obtains immediately
\[
\pfx{CGT}=\sum_V\, (\dim V)^\chi\, e^{-C_V \sum_f \lambda_f} ,
\]
where $\chi=c_2-c_1+c_0$ is the Euler characteristic of $M$ and the
sum runs as usual over the irreducible representations of $G$.
This agrees of course with the well known solution of lattice gauge
theory \cite{Mig:reclgt}.
Moreover, the diagrammatic solution carries over directly to the
quantum group case, yielding the same expression \cite{Oe:qlgt}.
(Then ``$\dim$'' denotes the quantum dimension and $C_V$ some
suitable quantum Casimir operator.)

%% file: figures/fig_hklim.tex
\[\rxy{,(0,-10);(0,10)**\dir{-}}
\quad\xleftarrow[\lambda\to 0]{}\quad
\rxy{
,(0,-10);(0,-3)**\dir{-}
,(0,3);(0,10)**\dir{-}
,(0,0)*\cir(3,0){}*{\lambda}
}\quad \xrightarrow[\lambda\to\infty]{}\quad
\rxy{
,(0,-10);(0,-4)**\dir{-}
,(0,4);(0,10)**\dir{-}
,(-3,-4);(3,-4)**\dir{-};(3,4)**\dir{-}
;(-3,4)**\dir{-};(-3,-4)**\dir{-}
}\]

%% file: figures/fig_rgflow.tex
$\rxy{,(0,0)*=(15,15){*};(50,0)*=(15,15){*}
="A"**\crv{(25,30)}?(0.5)*\dir{>}
,"A"**\crv{(25,10)}?(0.5)*\dir{>}
,"A"**\crv{(25,-10)}?(0.5)*\dir{>}
,"A"**\crv{(25,-30)}?(0.5)*\dir{>}
,(0,5)*{\text{FP}},(50,5)*{\text{FP}}
,(0,-5)*{\lambda=0},(50,-5)*{\lambda=\infty}
,(-7,0)*{\text{UV}},(57,0)*{\text{IR}}
}$

%% file: models2.tex
\section{Spin foam models of quantum gravity}
\label{sec:mod2}

In this section we consider certain spin foam models of Euclidean quantum
gravity in the light of
our approach to renormalization.
These are the Reisenberger model \cite{Rei:simpgr}, the
Barrett-Crane model \cite{BaCr:relsnet},
and a new model interpolating
between the latter and quantum BF theory.

For all considered models the problem of renormalization is nontrivial
as they depend on the discretization of space-time in a nontrivial
way.
Finding a solution to the renormalization problem being far beyond
the scope of this paper
we limit ourselves to a first step.
That is, we investigate the question of
amenability of these models to a renormalization treatment.
This implies identifying suitable local degrees of
freedom that can be acted upon by the renormalization groupoid.
Since these degrees of freedom are present neither in the Reisenberger
model nor in the Barrett-Crane model we propose modifications of both
models that carry such degrees of freedom.

\subsection{Motivation: BF theory plus constraint}

For the reader's convenience we review very briefly
the motivation of the present models as quantizations of general
relativity (see \cite{Bae:introsfoam} for more details).

Let $M$ be a compact four dimensional manifold (space-time) with a
$\Spin(4)$ principal bundle. Consider a 1-form $e$ (the vierbein) with
values in the Lie algebra of $\Spin(4)$ and a spin connection $A$. Let
$F$ be the curvature 2-form of $A$ and consider the action
\[
 S=\int_M \tr(e\wedge e\wedge F).
\]
This describes Euclidean general relativity (assuming $e$ to be
non-degenerate).

While this is difficult to quantize one can observe a certain
similarity with BF theory (\ref{eq:bfact}). Indeed we can take the BF
action if we introduce the additional constraint that the $B$-field
arises from an $e$-field, i.e.\
\begin{equation}
 B=e\wedge e .
\label{eq:constraint}
\end{equation}
The idea is now to quantize BF theory, which we know how to do
(see Section~\ref{sec:bf}) and impose the constraint (\ref{eq:constraint})
afterwards. The latter step is rather nontrivial and various proposals
for its implementation
have been made. Perhaps the best known ones are the model due to
Reisenberger and the one due to Barrett and Crane. These are the ones
we are going to discuss.

\subsection{The Reisenberger model}
\label{sec:Rei}

Since the spin group in four dimensions is $\Spin(4)=\SU(2)\times
\SU(2)$ we can decompose the spin bundle into two ``chiral''
components, one for each copy of $\SU(2)$. One can then
observe that one chiral component is enough to formulate classical
general relativity. This carries over to the above context of
obtaining general relativity by constraining BF theory.

The Reisenberger model \cite{Rei:simpgr} thus starts with quantum BF
theory 
(\ref{eq:bfpart}) with gauge group $\SU(2)$. It is formulated on a
simplicial decomposition of space-time.
The partition function is given by
\begin{equation}
 \pfx{Rei}=\int\prod_e \xd g_e \prod_v (z\sqrt{2\pi})^{-5}
 [e^{-\frac{1}{2 z^2} \Omega_{ij}\Omega^{ij}}]
\prod_f \delta(g_1 \cdots g_k) .
\label{eq:reipart}
\end{equation}
What is different as compared to quantum BF theory is the factor
\begin{equation}
 (z\sqrt{2\pi})^{-5}
 [e^{-\frac{1}{2 z^2} \Omega_{ij}\Omega^{ij}}]
\label{eq:Reifac}
\end{equation}
inserted for each vertex. $z$ is a real parameter of the model while
$\Omega_{ij}$ denotes a certain operator. $\Omega_{ij}$ is a sum of
operators each of which acts on a pair of delta functions associated
with the given vertex. Diagrammatically speaking it acts on all pairs
of wires (belonging to pairs of faces or 2-simplices) which are
associated with the vertex (i.e.\ which belong to the 4-simplex).
We will not give the full definition here as this is irrelevant for
our considerations.

The model as defined is not immediately amenable to
a treatment in our approach. A serious problem is the fact that it is
only defined for simplicial decompositions. Thus, although the
cellular moves can be applied in principle the resulting
configurations are not simplicial and thus not comparable to the
original model. An interesting alternative would be to try to define
the model for cellular decompositions through the moves. However this
would presumably involve solving the renormalization problem and is thus
beyond the scope of our current investigation.

Another prerequisite to a renormalization in our sense are
tunable local parameters. These are easily introduced into the
partition function. One can simply let the real parameter $z$ depend
on the vertex, i.e.\ introduce one $z_v$ per vertex.
What is more, for $z_v\to \infty$ we (essentially, i.e.\ up to
normalization) recover quantum BF theory. That is, our parameter space
contains a renormalization groupoid fixed point (up to an anomaly
which can be easily eliminated, see Section~\ref{sec:bf}).

\begin{figure}
\begin{center}
\input{figures/fig_rgfqg}
\end{center}
\caption{Qualitative picture of the renormalization groupoid flow for
  both the Reisenberger and the interpolating model. There is one
  ultraviolet fixed-point at $\lambda=0$ which is quantum BF
  theory. There is another special point at $\lambda=\infty$ 
  (or $z=0$) which for the interpolating model corresponds to the
  Barrett-Crane model.}
\label{fig:rgfqg}
\end{figure}
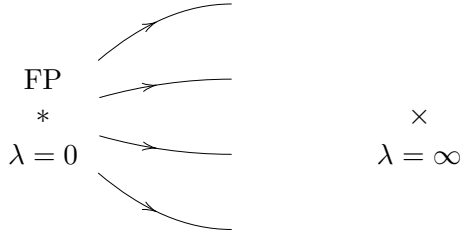

Assuming the difficulties can be overcome we can nevertheless make
some qualitative remark on the the renormalization groupoid flow. For
better comparability with the other models we consider
inverted parameters $\lambda\cong z^{-1}$. We also disregard the first
factor in (\ref{eq:Reifac}) which should not be relevant for the
qualitative picture. The model is rather similar
to cellular gauge theory in the vicinity of the quantum BF theory fixed
point. Indeed, in this region the arguments concerning the flow put
forward in Section~\ref{sec:rencgt} apply. That is, the flow points away
from the ``weak coupling limit''
$\lambda=0$ which is a repulsive ultraviolet fixed-point. On the
other hand, the point $\lambda=\infty$ is not a fixed point and it is
not clear how the flow behaves near it. Figure~\ref{fig:rgfqg} shows a
diagrammatic illustration of the situation.

\subsection{The Barrett-Crane model}
\label{sec:BaCr}

While different versions of the Euclidean Barrett-Crane model have
been proposed
\cite{DFKR:bcmod,PeRo:nobub,OrWi:gluingsimp,Pfe:dualbc,BCHT:sfriemqg}
we consider a version which appears to be the most natural one in our
framework.
For ease of terminology we refer to it in the following as ``the''
Barrett-Crane model.

While originally defined for simplicial complexes only it
was shown by Reisenberger that the Barrett-Crane model naturally
extends to more general complexes \cite{Rei:relvert}. Our treatment in
the following directly applies to general cellular decompositions.
It is closely related to the connection formulation of which an
extensive treatment can be found in \cite{Pfe:dualbc}.

We start with the full $\Spin(4)$ BF theory in
four dimensions. To describe the crucial step of implementing
the constraint (\ref{eq:constraint}) it will be convenient to
explicitly perform the decomposition into chiral components. Writing
each group element and matrix element
as a product of left-handed and right-handed chiral $\SU(2)$ component
the partition function (\ref{eq:bfpart}) decomposes into the product
of two BF partition functions
\begin{equation}
\pfx{BF}=\int\prod_e \xd g^L_e \xd g^R_e \prod_f
 \delta(g^L_1 \cdots g^L_k) \delta(g^R_1 \cdots g^R_k) .
\label{eq:chiralpart}
\end{equation}
Here the superscripts $L$ and $R$ denote the chiral components.

This step finds its diagrammatic
expression in the splitting of the circuit diagram for the partition
function (Figure~\ref{fig:circdiag}) into two identical diagrams for
$\SU(2)$, one for each chiral
component. Note that the representation labels for the diagrams are
now $\SU(2)$ representation labels, summed over independently.

We stick in the following to a purely diagrammatic representation of
the partition function. This avoids on the one hand writing
complicated formulas and on the other hand facilitates understanding
the structure of the model in view of the renormalization problem. It
also avoids extra complications arising for non-simplicial
decompositions in the spin foam formulation.

As a second step we replace each cable by a
sequence of two cables. This is a special case
of the (inverse) gauge fixing identity (Figure~\ref{fig:gaugefix}) and
thus does not change the
partition function. It implies algebraically another
(redundant) doubling of the group variables, so that for each chiral
component there is now one group variable attached to each end of
each edge.

\begin{figure}
\begin{center}
\includegraphics{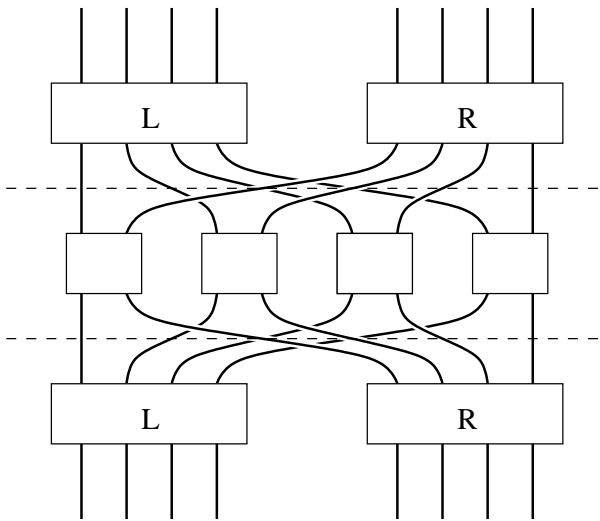}
\end{center}
\caption{In going from BF theory to the Barrett-Crane model one
inserts the cables between the dotted lines. There is one such cable
for each chiral pair of wires. Here is shown an edge with four wires,
the only case occurring in a simplicial decomposition in dimension
four. Labels and arrows on wires are omitted.}
\label{fig:bcproj}
\end{figure}

So far we have only rewritten the partition function without changing
it. The Barrett-Crane model is now 
obtained by inserting further cables on each edge between the cables
present. This is depicted in Figure~\ref{fig:bcproj}.
One cable is
inserted for each chiral pair of wires, so that the pair of wires goes
through the cable. This has the effect of projecting onto ``simple''
irreducible representations of $\Spin(4)$, i.e.\ those where both
chiral components are the same representation of $\SU(2)$. It is
designed to implement the constraint (\ref{eq:constraint}).
The partition function 
$\pfx{BC}$ of the Barrett-Crane model
is defined by this circuit diagram.
The conversion of the diagram into
a closed formula is again straightforward,
by inserting a group integral for each cable etc.

\begin{figure}
\begin{center}
\includegraphics{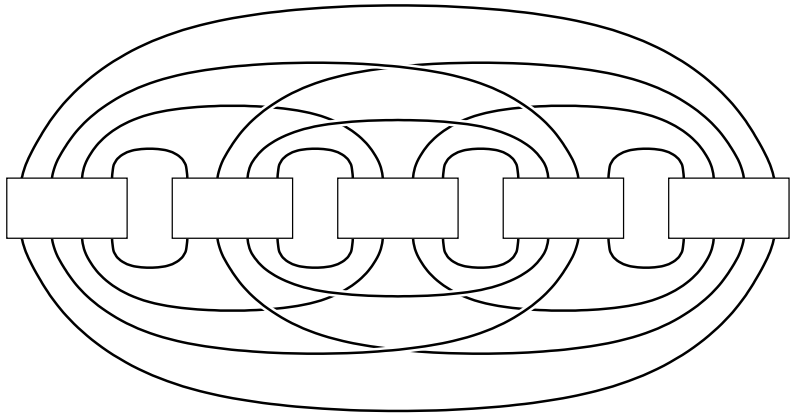}
\end{center}
\caption{The circuit diagram for a 4-simplex in the Barrett-Crane
model. Labels and arrows on wires are omitted.}
\label{fig:foursimp}
\end{figure}

Although the circuit diagram appears to be quite complicated at this
point it
can be simplified considerably. As the Barrett-Crane cables just carry
two wires with irreducible representations we can apply the tensor
product identity (Figure~\ref{fig:tpid}). This forces the two chiral
components to lie in the same representation and and leads to a
decomposition of the circuit diagram. Indeed we obtain a separate
circuit diagram for each vertex (or $n$-cell). After repeated
application of the gauge fixing identity (Figure~\ref{fig:gaugefix})
this consists of one cable per edge ($n-1$-cell) and one wire per face
($n-2$-cell) that meet the vertex ($n$-cell).
An easy way to
construct this diagram is as follows: Consider the part of the circuit
diagram for BF theory that lies in an $n$-cell. Cut it out (this cuts
cables in halves) and produce a mirror image. Now connect the diagram
and its mirror image in the obvious way to obtain a closed
diagram. The two parts correspond exactly to what were before the two
chiral components. For a 4-cell which is a simplex the diagram obtained
in this way is drawn in Figure~\ref{fig:foursimp}.

The whole partition function can be reexpressed as
\begin{equation}
\pfx{BC}=\sum_{V_f} (\prod_f (\dim V_f)^{2-|\partial f|})
 \prod_v A_v(V_f) .
\label{eq:bcpart}
\end{equation}
Here the summation is over labelings $V_f$ of faces with irreducible
representations of $\SU(2)$, $|\partial f|$ denotes the number of edges
of face $f$ and $A_v(V_f)$ is the value of the circuit diagram for the
vertex $v$ (with the appropriate labels).
The origin of the exponent $|\partial f|$ lies in the fact that each
application of the tensor product identity (Figure~\ref{fig:tpid})
produces a factor of inverse dimension.\footnote{Note that there is a
 subtlety associated with certain 
 cellular decompositions which does not arise for simplicial ones.
 Namely, there can be 2-cells which are not
 in the boundary of any 3-cell. In the dual language, these are faces
 which have no edge. In this case the associated wires (which are just
 loops) would not carry any projecting Barrett-Crane cable. That is,
 for those particular wires one still has to sum over chiral $\SU(2)$
 representations \emph{separately}. This would give a modification to
 (\ref{eq:bcpart}). On the other hand one could by
 definition restrict the chiral components to be equal also for those
 wires so as to arrive at (\ref{eq:bcpart}).
\label{fn}}

Other versions of the Barrett-Crane model agree with the present one
in the vertex amplitude $A_v(V_f)$, while differing in the choice of
weights for edges and faces.
For example, the Perez-Rovelli version \cite{PeRo:nobub} is obtained
if instead of
inserting one layer of Barrett-Crane cables as in
Figure~\ref{fig:bcproj} two such layers are inserted with another ``LR
layer'' inbetween.
We explain below why the present choice
stands out in view of a renormalization treatment.

As a first step in investigating renormalization properties of
the Barrett-Crane model we apply the moves to the model ``as is''.
One quickly sees that neither the $n$- nor
the $n-1$-move preserve the partition function in any obvious
way. Indeed, for the $n$-move one removes a subdivision between
$n$-cells (i.e.\ merges two vertices) so that before the move one has
two disconnected diagrams while afterwards there is one connected
diagram. Except for trivial cases there seems to be no way how they
could be equal. This is not a surprise. The relevant identity for
quantum BF theory was the gauge fixing identity. It reflects the
possibility of fixing a gauge and is based on gauge invariance (see
\cite{Oe:qlgt}). However, the $\Spin(4)$ gauge invariance is precisely
broken in the Barrett-Crane model (to a $\SU(2)$ subgroup) by the
introduction of the extra cables which act as projectors.

For the $n-2$-move the situation is similar to quantum BF
theory. Before the
move one has factors of dimension which are not present after the
move. See footnote~\ref{fn} however.

\subsection{An interpolating model}

As the Barrett-Crane model is not invariant under the cellular moves,
i.e.\ not topological, a renormalization treatment in our sense would
require local parameters. On top of that one would like to have a
renormalization groupoid fixed point in the ultraviolet
analogous to the weak coupling regime in cellular gauge theory
(Section~\ref{sec:ym}) and the large $z$ regime in the Reisenberger
model (Section~\ref{sec:Rei}).

\begin{figure}
\begin{center}
\includegraphics{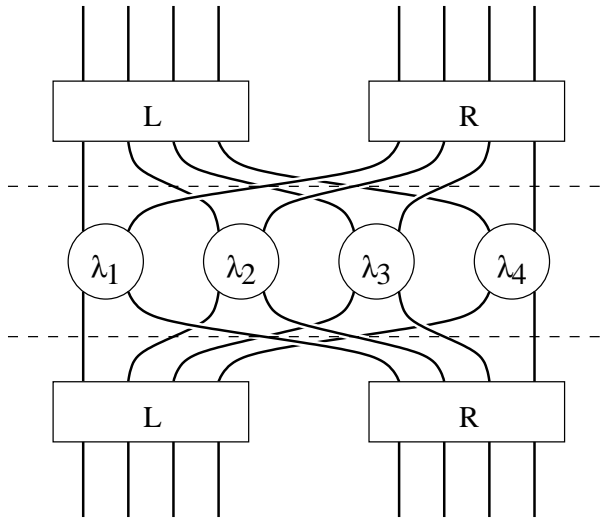}
\end{center}
\caption{Edge diagrams for the interpolating model. $\lambda_i$ are
positive real parameters. There is one such parameter per edge and
face. Labels and arrows on wires are omitted.}
\label{fig:newproj}
\end{figure}

Based on these requirements we propose a new model as follows:
The
Barrett-Crane model is defined as a modification of quantum BF theory
(which is topological). It is thus natural to \emph{tune}
the way this modification is performed. 
More precisely, as above, we start
with BF theory of $\Spin(4)$, rewriting it in terms of the chiral
decomposition and the extra doubling of the cables for each edge.
Then, instead of
proceeding to insert cables as in Figure~\ref{fig:bcproj} we insert
heat kernel factors as appearing in lattice gauge theory (see
Section~\ref{sec:ym}). Instead of arriving at the diagram of
Figure~\ref{fig:bcproj} we arrive at the diagram of
Figure~\ref{fig:newproj} per edge. This defines the partition function
$\pfx{Int}$ of the model.

The present definition uses the two limits of the heat kernel
operator $e^{-\lambda C}$ (Figure~\ref{fig:limhk}). For $\lambda\to 0$
we recover quantum BF theory while for $\lambda\to\infty$ we get the
usual Barrett-Crane model. The local parameters we have introduced are
one positive real number per edge \emph{and} face. If desired one could
decrease the number of parameters by requiring them to be the same
within a face or within an edge.

Considering qualitative aspects of the renormalization groupoid flow
we have now the same scenario as for the Reisenberger model
(Figure~\ref{fig:rgfqg}). At ``weak
coupling'' (small $\lambda$) the arguments of Section~\ref{sec:rencgt}
again apply. The renormalization groupoid flow is directed away from
the ultraviolet fixed point given by quantum BF theory at $\lambda=0$. 
On the other hand, much less is known about the point $\lambda=\infty$
which corresponds to the Barrett-Crane model.

However, a numerical study of different versions of the Barrett-Crane
model was conducted by Baez et al.\ \cite{BCHT:sfriemqg}. This
includes in particular the Perez-Rovelli version \cite{PeRo:nobub}. It
turns out that the partition
function in this case is strongly dominated by
contributions with almost all representations trivial.
It should be expected that our version of the Barrett-Crane model
behaves somewhat similar, although not quite as extreme.
This would suggest that at the point $\lambda=\infty$ of the interpolating
model (where the Barrett-Crane model resides)
we have ``almost'' a trivial theory, to be compared with the IR fixed
point in cellular gauge theory.

In retrospect we have now an additional argument for our choice of
weights for edges and faces for the Barrett-Crane model
(manifest in (\ref{eq:bcpart})). 
We see now that
it is not only natural in the way the model is constructed from
quantum BF theory but essential in obtaining an ultraviolet fixed
point in the parameter space of the interpolating model.

%% file: figures/fig_rgfqg.tex
$\rxy{,(0,0)*=(15,15){*};(50,0)*=(50,50){\times}
="A"**\crv{(25,30)}?(0.3)*\dir{>}
,"A"**\crv{(25,10)}?(0.3)*\dir{>}
,"A"**\crv{(25,-10)}?(0.3)*\dir{>}
,"A"**\crv{(25,-30)}?(0.3)*\dir{>}
,(0,5)*{\text{FP}}
,(0,-5)*{\lambda=0},(50,-5)*{\lambda=\infty}
}$

%% file: conclude.tex
\section{Discussion and conclusions}
\label{sec:conclude}

A crucial issue that we have not addressed in the present paper is
the role of observables. Indeed, for a sensible physical
interpretation what we
need is not a renormalization that keeps the partition function fixed
but one that preserves physical observables.
Nevertheless, the two might be closely related. Indeed, for
quantum BF theory the two notions coincide if
we take observables to be Wilson loops. This is why (quantum group
versions) of this model give rise both to invariants of manifolds and
knots (the knots being Wilson loops), see \cite{Oe:qlgt}.
The same is
true for cellular gauge theory (quantum Yang-Mills theory) in
dimension two. If we keep the ``area''
(the sum of the local parameters) inside a Wilson loop fixed, the
expectation value of the latter is exactly preserved by the
renormalization that we presented.

The situation is rather different for the models of quantum gravity
that we have considered. Here, even the question of what the correct
physical observables are is unresolved. Indeed, our qualitative
statements concerning the renormalization groupoid flow could be
(and probably will be) substantially altered if renormalization is
performed with respect to relevant observables.
A further point is that while
we have considered these models for given topology and cellular
decomposition it has been proposed to sum them over discretizations
(usually simplicial ones) of space-time and even over
topologies. However, to get control of such a sum it is presumably
still necessary to understand and relate the individual terms
(discretizations) and perform a renormalization in our sense.

One proposal for performing such a sum is that of a generating field
theory \cite{Bou:threedlat,Oog:toplat,DFKR:bcmod,ReRo:stfeyn}. This is
usually a field theory on the gauge group
which generates discretized space-times as its Feynman diagrams. This
is particularly easy to see when using our diagrammatic
language. Expressing propagators and vertices in terms of the diagrams
of Section~\ref{sec:circdiag}, a closed Feynman diagram becomes
exactly the circuit diagram representing the partition function of the
respective 
model for the corresponding discretization.\footnote{Strictly
speaking, one has to view Feynman diagrams as intertwiners in
the representation category of the group. This is explained (among
other things) in
\cite{Oe:bqft,Oe:introbqft}. Then one uses the circuit diagrams to
represent these intertwiners \cite{Oe:qlgt}.}

In general, a renormalization with respect to observables is a
weaker problem than a local renormalization with respect to the
partition function. Namely, the latter implies the former but not the
other way round. For example, in the quantum BF theory case, the
anomaly cancels in expectation values. A related issue is that asking
for an exact renormalization is often too much. A coarse
discretization is not supposed to reproduce all the physics of a finer
one. Thus, one usually requires a renormalization to preserve
observables only in some approximate sense, for example have them
converge in an ``infinite refinement limit'' as in lattice gauge
theory.

While we have limited ourselves here to closed manifolds
our framework extends straightforwardly to manifolds with
boundary. Indeed, our definitions in Sections~\ref{sec:cellular} and
\ref{sec:renorm} carry over practically unchanged.
In addition, gluing along boundaries
is essentially straightforward for the models we consider. For
quantum BF theory and cellular gauge theory this is
already explained in \cite{Oe:qlgt}. In a setting with boundaries
renormalization groupoid fixed points correspond to topological
quantum field theories (at least if an exact renormalization of the
partition function is required).

One could also envisage a special treatment of boundaries where they
carry extra parameters. This might be desirable if a fixed background
is attached to a boundary. Such a situation could occur for example if
in quantum gravity one wishes to put quasi-classical or coherent states
on the boundary.

As we have emphasized it is rather important for renormalization in
our sense to work to allow for cellular decompositions and not
restrict to simplicial ones. This is a problem in the Reisenberger
model (considered in Section~\ref{sec:Rei}) whose definition strongly
uses the fact that cells are simplices. A generalization of the model
in this direction is thus desirable. In contrast, the formulation
of the Barrett-Crane model (considered in Section~\ref{sec:BaCr}) in
terms of cellular decompositions is straightforward. In this case
however,
local parameters allowing for a renormalization groupoid flow are
absent. This prompted us to define an interpolating model that
contains both the Barrett-Crane model and quantum BF theory in its
(local) parameter space. To define this we ``imported'' the idea of
heat kernel factors from lattice gauge theory. This allows a
continuous switching between a ``weak coupling limit'' of quantum BF
theory and the ``strong coupling limit'' of the Barrett-Crane model.

We suggested that the renormalization groupoid flow both in the
Reisenberger and in the interpolating model should behave similarly as
for cellular gauge theory in the region of ``weak coupling'' (small
$\lambda$). All these models share the same renormalization
groupoid fixed point of quantum BF theory in the ultraviolet. On the
other hand, while having a special point in the ``strong coupling
limit'' of $\lambda=\infty$ this is an infrared fixed point only for
cellular gauge theory. Nevertheless, both for the Reisenberger model
and for the interpolating model one should expect behaviour somewhat
similar to cellular gauge theory also in this region. In both cases,
there are projection operators which are ``maximally turned on'' at
$\lambda=\infty$ comparable to the heat kernel operators
in cellular gauge theory. In contrast to the latter they do not lead
to a complete restriction to trivial representations. Nevertheless,
they should have an effect approaching this.

For the Barrett-Crane model
numerical investigations
by Baez et al.\ seem to indicate indeed that the partition function is
strongly dominated by terms where almost all representations are
trivial \cite{BCHT:sfriemqg}. Baez et al.\ suggest that such a
situation might indicate a bad
choice of weights for edges and faces.
Our conclusion is rather different. We propose that the
Barrett-Crane model should be considered as just a point in the
parameter space of a more general model (the interpolating model)
which is rather more amenable to a renormalization. From this point of
view it is even a welcome feature if the Barrett-Crane partition
function is strongly dominated by terms with almost all
representations trivial.
This would give the interpolating model at ``strong coupling'' a
region in parameter space which ``almost'' contains an infrared fixed
point of the renormalization groupoid.

Markopoulou has made the interesting proposal
\cite{Mar:alggrain,Mar:coarsesf} to
apply the renormalization methods of perturbative quantum field theory
to spin foam models.
Based on the structural
similarity of spin foams with Feynman diagrams (which becomes a strict
correspondence in the generating field theory approach mentioned
above) she suggests a Bogoliubov type recursion equation for spin
foams (formulated in a Hopf algebraic language).
Her concept of coarsening is strictly based on spin foams but
essentially coincides with the one coming from cellular
decompositions in this context. Thus, one might hope that
her approach can fruitfully complement the one presented
here. In particular, Markopoulou's approach might be useful in
eliminating certain infinities (as the corresponding techniques in
perturbative quantum field theory do).
Such infinities arise for example from
summing over infinitely many irreducible representations, as in
quantum BF theory, compare (\ref{eq:bfexp}).

The basic ideas on renormalization introduced in
Sections~\ref{sec:cellular} and \ref{sec:renorm} are not specific to
the gauge type models that we have considered in the later
sections. Indeed they should
be applicable to a wide range of models (including spin systems for
example). Furthermore, while we have emphasized the role of background
independence, this is not a requirement. Indeed, the proposed methods
might be applicable to situations where a metric background is
present, but the discretization of space-time is nevertheless
irregular (as in ``random geometry'' models). In such a situation
metric information might be converted into local parameters. Indeed,
we saw that this was a possible point of view in our cellular
gauge theory example of Section~\ref{sec:ym}.

Much of our treatment seems to depend on Conjecture~\ref{con:cellrel}
and indeed it would be important to establish its validity (or
non-validity). However, the situation is rather less serious. Firstly,
even if the conjecture was false in its present form, it is very
likely to hold under some mild technical assumptions about the
manifold or cellular decomposition (such as piecewise linearity).
Such assumptions very likely would have little or no physical
relevance. Secondly, for the application to a given model it is
usually not necessary that the conjecture holds in all its
aspects. For example, for all the models considered in this paper only
the cells of dimension $n$, $n-1$, and $n-2$ are relevant. Thus, a
much weaker form of the conjecture suffices.
Thirdly, considering a certain subclass of refinements or coarsenings
might be enough for a sensible renormalization in a given
context. Again, this would require only a weaker form of the
conjecture. 

While we worked throughout with a topological manifold it should be
possible to do away even with this ``background''. A suitable setting
in the present context would probably be that of (finite)
CW-complexes. Conjecture~\ref{con:cellrel} is definitely false if
``$n$-dimensional manifold'' is replaced by ``$n$-dimensional
CW-complex''. This is not surprising as the complex can contain
``pieces'' of lower dimension. On the other hand
one could consider all cellular
moves for all dimensions up to $n$ (the dimension of the cell of
maximal dimension). This would give $n(n+1)/2$ moves. The analogue of
Conjecture~\ref{con:cellrel} would then state something like that any
two finite 
decompositions as CW-complexes (if they exist) of the same topological
space are related by a sequence of these moves. The basic
considerations of Section~\ref{sec:renorm} concerning renormalization
could then be carried over practically unchanged. One could be even
more general by considering combinatorial complexes. However,
a quantum group generalization of models as considered here would be
definitely lost in this
case as it depends on certain topological information \cite{Oe:qlgt}.

%% file: acknowledge.tex
\section*{Acknowledgements}

I would like to thank Hendryk Pfeiffer and Carlo Rovelli for valuable
comments on the manuscript. Furthermore, 
I would like to thank Thomas
Krajewski and Valentin Zagrebnov for discussions.
This research was supported by the
European Union through a Marie Curie Fellowship grant.